\pdfoutput=1
\documentclass{vldb}

\usepackage{graphicx}
\usepackage{balance}  %
\usepackage[format=plain,labelfont=bf]{caption,subcaption}

\usepackage{multicol} %
\usepackage{tabularx}
\usepackage{array}
\usepackage{booktabs}
\usepackage{rotating}
\usepackage{url}
\usepackage{enumitem}
\usepackage{float}
\usepackage{placeins}
\usepackage{fancyvrb}
\usepackage{listings}
\usepackage{bbm}
\usepackage{xspace}

\usepackage{color, colortbl}
\definecolor{Gray}{gray}{0.9}

\usepackage{tikz}
\usepackage{tikz-cd}

\usepackage{amsmath}
\newcommand{\ra}[1]{\renewcommand{\arraystretch}{#1}}

\newcommand\numberthis{\addtocounter{equation}{1}\tag{\theequation}}

\usepackage{amsthm}
\theoremstyle{definition}
\newtheorem{definition}{Definition}[section]

\DeclareMathOperator*{\argmin}{arg\,min}

\newcommand{\stitle}[1]{\vspace{0.5ex}\noindent{\bf #1}}
\newcommand{\eat}[1]{}

\usepackage{xcolor}

\newcommand{\pari}[1]{{{\color{blue}(pari: #1)}}}

\newcommand{\Y}{Y^{true}}
\newcommand{\hatY}{Y^{est}}

\def\FlowLoss{Flow-Loss\xspace}
\def\Priority{Prioritized Q-Error\xspace}
\def\PostgresError{Postgres Plan Cost\xspace}

\def\OurDataset{Cardinality Estimation Benchmark\xspace}
\def\CEB{CEB\xspace}
\def\PlanGraph{Plan graph\xspace}
\def\plangraph{plan graph\xspace}
\def\PPE{PPC\xspace}

\vldbTitle{Flow-loss: Learning Cardinality Estimates that Matter}

\vldbDOI{https://doi.org/10.14778/xxxxxxx.xxxxxxx}
\vldbVolume{12}
\vldbNumber{xxx}
\vldbYear{2020}

\begin{document}

\title{Flow-Loss: Learning Cardinality Estimates That Matter}

\author{
 \alignauthor
 Parimarjan Negi$^{1}$,
 Ryan Marcus$^{12}$,
 Andreas Kipf$^{1}$,
 Hongzi Mao$^{1}$,\\
 Nesime Tatbul$^{12}$,
 Tim Kraska$^{1}$,
 Mohammad Alizadeh$^{1}$\\
 \affaddr{$^1$MIT CSAIL \quad $^2$Intel Labs}
 \email{\{pnegi, rcmarcus, kipf, hongzi, tatbul, kraska, alizadeh\}@mit.edu}
}

\maketitle

\begin{abstract}
Previous approaches to learned cardinality estimation have focused on improving average estimation error, but not all estimates matter equally. Since learned models inevitably make mistakes, the goal should be to improve the estimates that make the biggest difference to an optimizer.
We introduce a new loss function, \FlowLoss, that explicitly optimizes for better query plans by approximating the optimizer's cost model and dynamic programming search algorithm with analytical functions. At the heart of \FlowLoss is a reduction of query optimization to a flow routing problem on a certain plan graph in which paths correspond to different query plans. To evaluate our approach, we introduce the \OurDataset (\CEB) which contains the ground truth cardinalities for sub-plans of over $16K$ queries from $21$ templates with up to $15$ joins.
We show that across different
architectures and databases, a model trained with \FlowLoss improves the cost of plans (using the PostgreSQL cost model) and query runtimes despite having worse estimation accuracy than a model trained with Q-Error. When the test set queries closely match the training queries, both models improve performance significantly over PostgreSQL and are close to the optimal performance (using true cardinalities). However, the Q-Error trained model degrades significantly when evaluated on queries that are slightly different (e.g., similar but not identical query templates), while the \FlowLoss trained model generalizes better to such situations. For example, the \FlowLoss model achieves up to $1.5\times$ better runtimes on unseen templates compared to the Q-Error model, despite leveraging the same model architecture and training data.

\if 0
We find that when the test set queries closely match the training distribution, both the models improve performance significantly over estimates from PostgreSQL, and are close to the optimal performance (if we gave PostgreSQL true cardinalities).
Notably, the \FlowLoss optimized model also does relatively well when the test set queries are slightly different, being close to optimal even on the Join Order Benchmark (JOB) despite being trained on the \OurDataset.
Meanwhile, the model optimizing for Q-Error degrades significantly.
In terms of query runtimes, the \FlowLoss model is almost $3x$ better on JOB, and up to $1.5x$ better on unseen templates compared to the Q-Error baseline, despite leveraging the same model architecture and training data. \pari{fix the result statistics based on latest results}
\fi

\end{abstract}

\section{Introduction}
\label{sec:intro}

\begin{figure}[t]
    \centering
    \includegraphics[width=1.0\columnwidth]{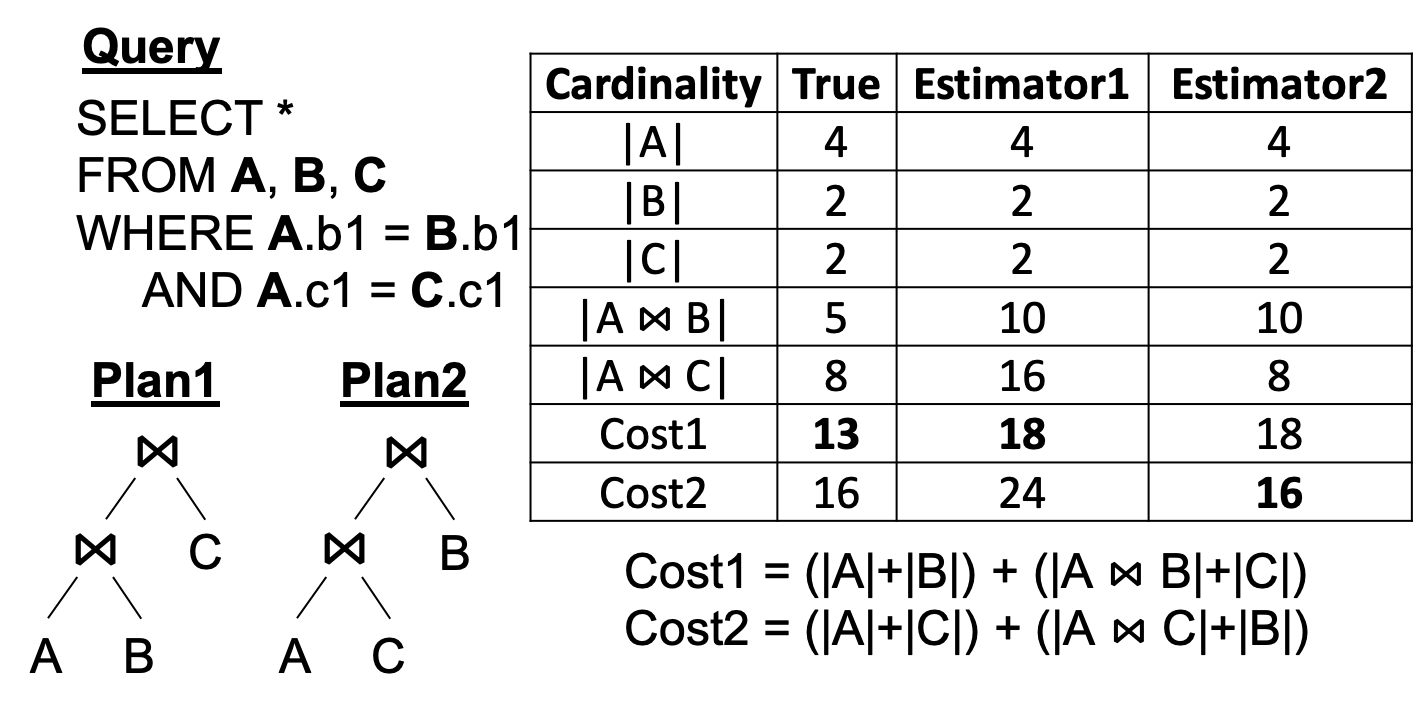}
    \caption{For this example, we use the sum of the cardinalities as the cost of a plan. With true cardinality values, Plan1 is cheaper than Plan2. This is also the case with Estimator1. Interestingly, however, although Estimator2’s cardinality values have smaller error than those of Estimator1, they will mislead the optimizer to choose Plan2.}
    \label{fig:eg1}
\end{figure}

Cardinality estimation is a core task in query optimization for predicting the sizes of \emph{sub-plans}, which are intermediate operator trees needed during query optimization.
Query optimizers use these estimates to compare alternative query plans according to a cost model and find the cheapest plan.
Recently, machine learning approaches to cardinality estimation have been successful in improving estimation accuracy~\cite{mscn,localmodels,lightweightmodels,deepdb,naru}, but they largely neglect the impact of improved estimates on the generated query plans.
This is the first work (known to us) that learns cardinality estimates 
by directly optimizing for the cost of query plans generated by an optimizer.

All learned models will have non-trivial estimation errors due to limitations in model capacity, featurization, training data, and differences between  training and testing conditions (e.g., due to changing workloads). Therefore it is crucial to understand which errors are more acceptable for the optimizer.
Unsupervised models learn from the data --- but they will use model capacity for sub-plans that never occur since they treat every potential query as equally likely.
Supervised models require representative workloads, but learn more efficiently by focusing model capacity on likely sub-plans.
However, all estimates are not equally important.
While an optimizer's decisions may be very sensitive to estimates for some sub-plans (e.g. join of two large tables), other estimates may have no impact on its decisions.

As a drop-in replacement for the well known Q-Error~\cite{qerror} loss function used to train supervised cardinality estimation models, we propose \emph{\FlowLoss}, a loss function that explicitly emphasizes estimates that matter to query performance for a given workload. \FlowLoss takes the idea of focusing model capacity to its logical extreme --- encouraging better estimates only if they improve resulting query plans.
For instance, consider Figure \ref{fig:eg1}: Estimator2 corrects Estimator1's estimate of $A \Join C$, but it actually leads to a worse plan (Plan 2), because the  relative cardinalities ($A \Join B$ vs. $A \Join C$) are incorrect. A loss function using \FlowLoss will show no error for Estimator1, while nudging Estimator2 to correct the relative cardinalities of these two joins. 

At its core, \FlowLoss computes the gradient of the cost of a query plan w.r.t. the cardinality estimates used to generate the plan.
To do this, we assume a simplified cost model 
and recast the dynamic programming (DP) algorithm for query optimization as a shortest path problem, which we approximate with a smooth and differentiable analytical objective function.
This lets us use gradient descent based techniques to improve the estimates that are most relevant to improving the query plans.
We show that improving cardinalities w.r.t. this objective also improves the quality of plans of the complex PostgreSQL cost model and optimizer.

There are two main benefits of training models to minimize \FlowLoss. First, similar to how attention-based models in natural language processing~\cite{attention} treat certain parts of the input as more important than others, \FlowLoss highlights which sub-plans are most relevant to the query optimizer. This helps a model focus its limited capacity on robustly estimating the sizes of such sub-plans. Across various scenarios, we show that \FlowLoss trained model have worse average estimation accuracy than Q-Error trained models, but improve the cost of generated plans. For instance, we show in an ablation study that models trained with \FlowLoss can adapt to removing various components of the featurization scheme, and still do equally well.
Meanwhile, ablations cause the Q-Error models to get up to $2\times$ worse w.r.t. PostgreSQL costs.
Second, by having a larger tolerance for errors on less critical sub-plans, training with \FlowLoss can avoid overfitting the model to cardinalities for which precise estimates are not needed, thereby leading to simpler models without sacrificing query performance.
Such simpler models typically generalize better. We show that models trained using Q-Error can be brittle, and can lead to significant regressions when the query workload diverges slightly from the training queries; for example, achieving up to 1.5$\times$ slower average runtimes than the default estimates in PostgreSQL, despite better overall estimation accuracy. Meanwhile, models trained using \FlowLoss do not see such drops in performance, and typically improve over PostgreSQL estimates, even across different workloads than those used in training. 

Our key contributions are:
\begin{itemize}[noitemsep,topsep=0pt,leftmargin=5mm,itemsep=0mm,itemindent=0mm]
    \item \stitle{\PostgresError (\PPE).} Based on Moerkotte et al.'s~\cite{qerror} cost model-based proxy for the runtime of a query plan, we introduce \PPE as a metric to evaluate the goodness of cardinality estimates in terms of their impact on query optimization. We show that it corresponds closely to runtimes, and provide an implementation to easily evaluate the performance of cardinality estimation models on \PPE using PostgreSQL. 
    \item \stitle{\FlowLoss.} We introduce \FlowLoss, a smooth and differentiable approximation of \PostgresError which can be optimized by any supervised learning model with gradient descent. 
    \item \stitle{\OurDataset (\CEB).} 
    We create a new tool to generate challenging queries based on templates in a semi-automated way. We use this to create the \OurDataset, which is over $100 \times$ larger than the Join Order Benchmark (JOB)~\cite{qoleis}, and has more complex queries.
    \item \stitle{Training with AQP estimates.} A challenge for using supervised cardinality estimation models in practice is that collecting ground truth data is expensive.
    However, precise estimates are not needed for near-optimal plans.
    We show that almost equally good query plans can be generated using  models trained with \FlowLoss on data collected using approximate query processing (AQP), which is $10-100 \times$ faster than computing true values. Since Q-Error tends to overfit, it is less robust to noisy training data generated via AQP. In fact, when using AQP training data, at the $99th$ percentile, models trained with Q-Error have over $10\times$ worse Q-Error and PostgreSQL costs than models trained with \FlowLoss.
\end{itemize}

\section{Related Work}
For cardinality estimation, traditional approaches have used histograms~\cite{stholes}, sampling~\cite{ibjs}, kernel density estimation~\cite{kde}, wavelets~\cite{matias1998wavelet}, or singular value decomposition~\cite{poosala1997selectivity}.
Recently, machine learning approaches have shown high estimation accuracy.
Many works focus on single-table selectivity estimates~\cite{quicksel,naru,made,lightweightmodels}, but while this is useful in other contexts, such as approximate query processing, it is non-trivial to extend such models to joins using join sampling~\cite{joinsampling}.
Learned cardinality estimation for joins can be categorized into \emph{unsupervised} (data-driven, independent of query workload) and \emph{supervised} (query-driven) approaches. Unsupervised approaches for cardinality estimation include Probabilistic Graphical Models~\cite{pgmselectivity,graphicalselectivity}, Sum-Product Networks~\cite{deepdb}, or deep autoregressive models~\cite{neurocard}. %
NeuroCard~\cite{neurocard} is the most advanced of these approaches, but it still does not support the complex analytical workloads studied in this work (e.g., queries with self joins).
That being said, any unsupervised model can be integrated into our approach by providing their estimates as features. 

Supervised approaches use queries with their true cardinalities as training data to build a regression model.
Our work builds on the approach pioneered by Kipf et al.~\cite{mscn}, which uses a single deep learning model for the whole workload. While several supervised learning-based works report improved estimation accuracy~\cite{staterepresentations,mscn,localmodels,cardinalitysharedclouds,lightweightmodels,lightweightmodels2}, only a few actually demonstrate improved query performance~\cite{ivanov2017adaptive,deepcardinalitysurvey,costguided}.
Our approach seeks to learn the cardinalities used by a traditional DBMS optimizer, while using the optimizer's search and cost algorithms for query optimization. 
Recently, there have been several other learning approaches to improve query performance which are complementary to our methods: learning the complete optimizer~\cite{handsfree,neo,joinsdeep18}, learning to use the optimizer's hints~\cite{bao}, learning the cost model~\cite{learnedcosts}, re-optimization~\cite{perron2019learned, skinnerdb}, and bounding worst case cardinalities to avoid bad plans~\cite{cai2019pessimistic}.

\begin{figure}[t]
\includegraphics[width=\columnwidth]{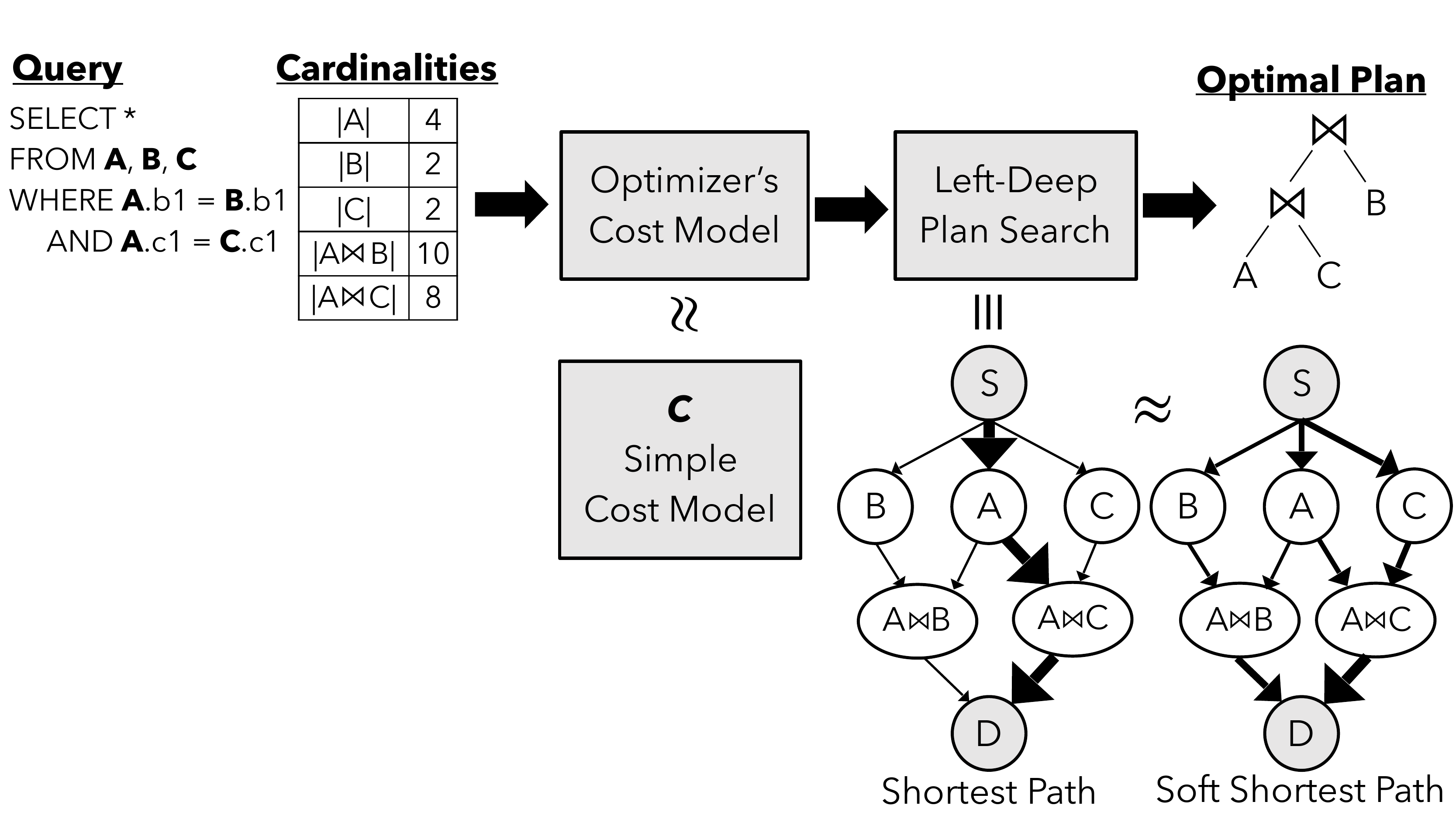}
\caption{The query optimization process has two non-differentiable components: the cost model and the plan search algorithm. We develop differentiable approximations for these so we can understand how sensitive query plans are to changes in cardinality estimates. }
\label{fig:flow_loss_overview}
\end{figure}

\section{Overview}
\label{sec:key_idea}
In this section, we will provide the high-level intuition behind our approach, which will be formalized in the next sections.
We target supervised learning methods that use a parametric model, such as a neural network, to estimate cardinalities for sub-plans required to optimize a given query.
Today, such models are trained using loss functions that compare true and estimated cardinalities for a given sub-plan, such as Q-Error.
\begin{definition}
\label{def:qerror}
Q-Error.
\begin{align}
    \mbox{Q-Error} (y^{true}, y^{est}) = \max(\frac{y^{true}}{y^{est}}, \frac{y^{est}}{y^{true}}).
\end{align}
\end{definition}

Such a loss function treats every estimate as equally important. Instead, we want a loss function that will focus model capacity on improving accuracy of estimates that matter most to the quality of the plans produced by the optimizer, while tolerating larger errors for other estimates. This loss function will need to be differentiable so we can optimize it using standard gradient descent methods.

To understand how cardinality estimates impact the resulting query plan, let us consider the basic structure of a query optimizer. There are two independent components, as highlighted in Figure \ref{fig:flow_loss_overview}: {\em (i)} a {\em cost model}, which outputs a cost for every join given the cardinality estimates for all sub-plans. {\em (ii)} a {\em DP search algorithm}, which finds the cheapest query plan. Our goal is to approximate both components using analytical functions that can be combined into a single, differentiable loss function:
\begin{equation}
    \hatY \xrightarrow{C(\cdot)} \text{Join-Cost} \xrightarrow{S(\cdot)} \text{Plan}.
\end{equation}
Here $C(\cdot)$ maps the cardinality estimates, $\hat{Y}$, to the cost of each join, and $S(\cdot)$
 maps the join costs to the optimal plan.  
 
Approximating the cost model as an analytical function is straightforward since it is already represented using analytical expressions. In principle, we can make this function as precise as we want, but we found that a very simple approximation with terms to cost joins with or without indexes works well in our workloads (Definition \ref{def:cost_model}). 

However, the DP search algorithm is non-trivial to model analytically. Our key contribution is in developing a differentiable analytical function to approximate {\em left-deep} plan search. Left-deep plans join a single table to a sub-plan at each step. Our construction exploits a connection between left-deep plan search and the shortest path problem on a certain ``plan graph''. While we focus on left-deep search for tractability, the resulting loss function improves the performance for all query plans, as the sub-plans required for costing left-deep plans are the same as required for all plans.

Figure~\ref{fig:flow_loss_overview} shows the plan graph corresponding to a simple query that joins three tables $A$, $B$, and $C$. Every edge in the plan graph represents a join and has a cost, and every path between two special nodes, $S$ and $D$, represents a left-deep plan. The DP search algorithm outputs the cheapest plan, i.e. the shortest path. When cardinality estimates change, they change the cost of the edges in the plan graph, possibly changing the shortest path. Therefore, to capture the influence of cardinality estimates on the plan analytically, we need an expression to relate edge costs to the shortest path in the plan graph. 

But this alone is not enough. The shortest path is insensitive to small changes to most edge costs (and hence, small changes to most cardinality estimates). For instance, consider any edge not on the shortest path; slightly increasing or decreasing the cost of that edge would not change the shortest path. Therefore an analytical function based on the shortest path would not have a {\em gradient} with respect to the cost of such edges. This would make it impossible for gradient-descent-based learning approaches to improve.

We tackle these challenges by using a soft approximation to the shortest path problem. In this formulation, the plan graph is viewed as an electrical circuit, with each edge having a resistance equal to its cost. One unit of current is sent from $S$ to $D$, split across paths in a way that minimizes the total energy consumed.\footnote{Electrical flows have been used to construct the most efficient polynomial time algorithms for approximating the closely related maximum flow problem in graphs~\cite{maxflow1,maxflow2,maxflow_madry}.} This formulation has two advantages over shortest path. First, it provides an explicit, closed-form expression relating the edge resistances (costs) to the amount of current on every path. Second, it does not suffer from the non-existent gradient problem described above. In an electrical circuit, the current is not exclusively sent on the path with the least resistance (i.e., the path corresponding to the cheapest plan). Instead, all low-resistance paths carry a non-negligible amount of current. Therefore, changing the resistance (cost) of an edge on any of these paths will affect the distribution of current across the entire circuit. The implication in our context is that all joins involved in low-cost query plans matter (even if they do not appear in the cheapest plan). This aligns with the intuition that the optimizer is sensitive to precisely these joins: changing their cost could easily change the plan it picks.

\if 0

\fi
\section{Definitions}
\label{sec:definitions}

\begin{figure}[t]
\centering
\includegraphics[width=\linewidth]{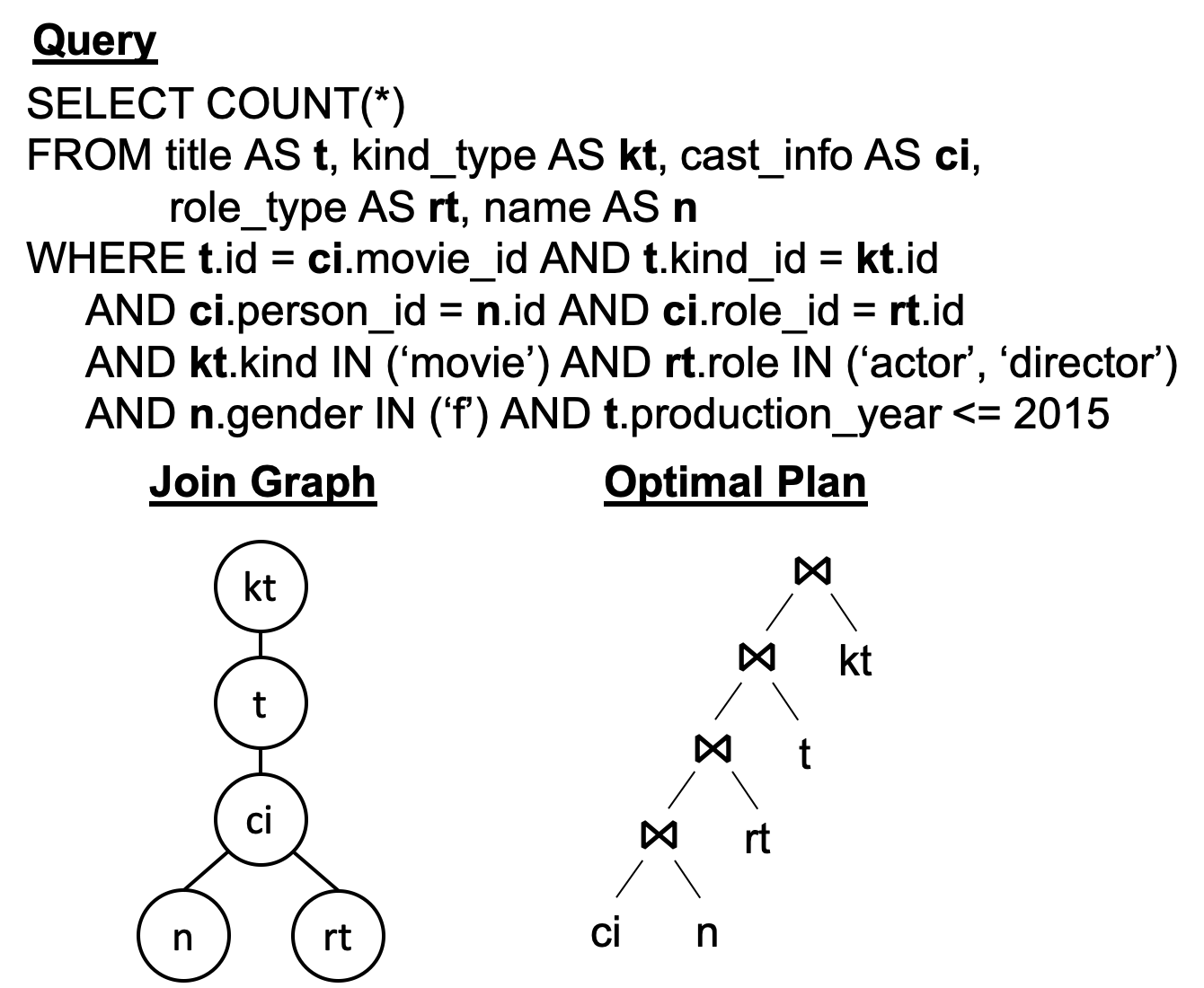}
\caption{Join graph and optimal plan for sample query $Q_1$ on the IMDb database.}
\label{fig:q1_figure}
\end{figure}

\begin{figure}[t]
    \centering 
    \includegraphics[width=1.0\linewidth]{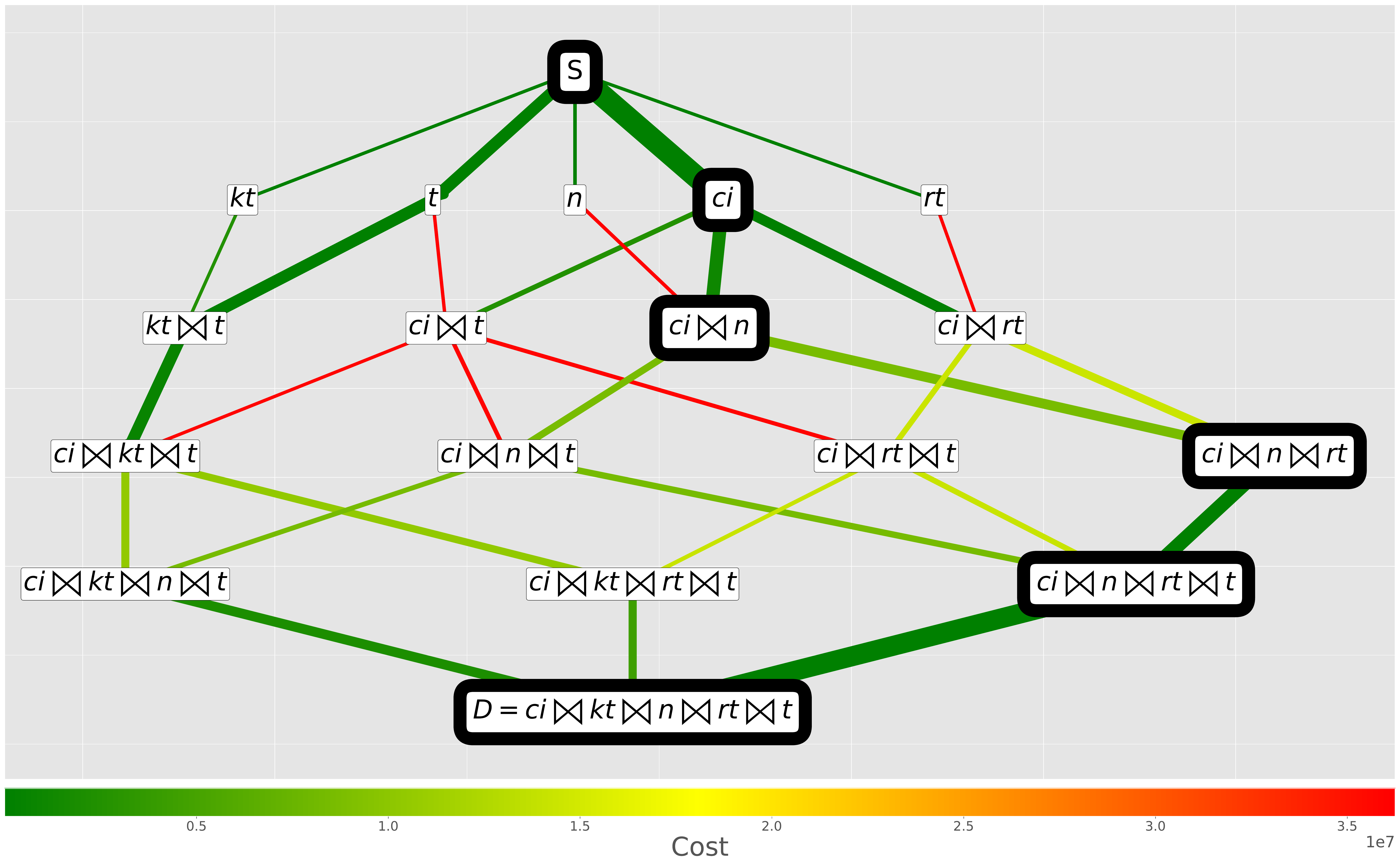}
    \caption[]%
    {{ \PlanGraph (Definition \ref{def:subquery_graph}) for query $Q_1$. The cheapest path,  P-Opt($\Y$), is highlighted. The edges are colored according to $C(e,\Y)$. The relative thickness of the edges represent the flows computed by Equation \ref{eq:flows_optimization}, F-Opt$(\Y)$.
    }}    
    \label{fig:subquery_graph}
\end{figure}

This section formally defines the plan graph and the concepts we use to develop our new loss function, \FlowLoss. 
As a running example, we will consider the query \textit{$Q_1$} (Figure \ref{fig:q1_figure}) on the Internet Movie Database (IMDb). Throughout this work, joins refer to inner joins, and we ignore cross-joins. For simplicity, we assume all joined columns have an index.

\begin{definition} {\em Sub-plan.} Given query $Q$, a sub-plan is a subset of tables in $Q$ that can be joined using inner joins. In query $Q_1$ (cf. Figure \ref{fig:q1_figure}), $kt \Join t$ is a sub-plan but $kt \Join ci$ is not.
\end{definition}

\begin{definition}
\label{def:subquery_graph}
{\em \PlanGraph.} Given query $Q$, the \plangraph is a directed acyclic graph (V,E) where $V$ is the set of all sub-plans, and there is an edge corresponding to every join in $Q$ between a sub-plan and a base table, i.e. $(u,v) \in E$ if and only if $v = u \Join b$ for a base table $b$. 
For convenience, we add a node $S$ for the empty set, which has an edge to all nodes containing exactly one table. We use $D$ to denote the node consisting of all tables.  Figure~\ref{fig:subquery_graph} shows the plan graph for query $Q_1$.

\end{definition}

\begin{definition}
{\em Path / Plan}, $P$.  A path (sequence of edges) from $S$ to $D$ in the \plangraph. Any left-deep plan corresponds to a path from $S$ to $D$. 
For instance, the plan $(((t \Join kt) \Join ci) \Join n) \Join rt$ for query $Q_1$ corresponds to: $S \rightarrow t \rightarrow t \Join kt \rightarrow  t \Join kt \Join ci \rightarrow  t \Join kt \Join ci \Join n \rightarrow  D$ in Figure \ref{fig:subquery_graph}.

\end{definition}

\begin{definition}
{\em Cardinality vector $Y$.} The cardinalities for each node (sub-plan) in the \plangraph. We use $\Y$ and $\hatY$ to refer to true and estimated cardinalities. 

\end{definition}

\begin{definition}
\label{def:cost_model}
$C(e,Y).$ A cost model which takes as input an edge (join) $e$ in the \plangraph and assigns it a cost given the cardinality vector $Y$. 
In this paper we use the following simple cost model:
\begin{align*}
C( (u,v), Y) = \min (|u| + \lambda |b|, |u| \cdot |b|)
\numberthis
\label{eq:phi}
\end{align*}
where $b$ is a base table s.t. $u \Join b = v$ and $|u|, |b|$ are cardinalities of $u$ and $b$ given by $Y$. The term $|u| \cdot |b|$ models nested loop joins without an index, and $\lambda=0.001$ is used to model an index on $b$. Figure \ref{fig:subquery_graph} shows the cost of each edge in query $Q_1$.
\FlowLoss can use a more precise cost model (e.g., with terms for other join operators such as hash join), but we found this simple model is effective in our workloads. 
\S \ref{sec:fla_cm} analyzes how well it approximates the PostgreSQL cost model. 
\end{definition}

\begin{definition}
$\text{P-Opt(Y).}$ The cheapest path (plan) in the plan graph with edge costs given by $C(e, Y)$:
\begin{align}
    \label{eq:opt_plan}
    \text{P-Opt}(Y) = \argmin_{P} \sum_{e \in P} C(e, Y).
    \numberthis
\end{align}
For example, given $\Y$, the cheapest path $\text{P-Opt}(\Y)$ is highlighted in Figure \ref{fig:subquery_graph}. %
We will use the terms ``cheapest'' and ``shortest'' path interchangeably.

\end{definition}

\begin{definition}
\label{def:plan_cost}
$\text{P-Cost}(\hatY, \Y).$ The true cost of the optimal path (plan) chosen based on cardinality vector $\hatY$: 
\begin{equation}
    \label{eq:plan_cost}
    \text{P-Cost}(\hatY, \Y) = \sum_{e \in \text{P-Opt}(\hatY)}  C(e, \Y).
\end{equation}
P-Cost can be viewed as an alternative to loss functions like Q-Error to compare estimated and true cardinalities $\hatY$ and $\Y$. It finds the cheapest path using $\hatY$, i.e. P-Opt$(\hatY)$, and then sums the {\em true} costs of the edges in this path using $\Y$. Note that for a fixed $\Y$, P-Cost takes its lowest value when $\hatY = \Y$. 
\end{definition}

\noindent{\bf Remark.} As defined, P-Cost is not a distance metric~\cite{distance-metric} (e.g., it does not satisfy the symmetry property). However, this does not affect its use in our loss function. In an online appendix~\cite{online_appendix}, we use P-Cost to construct a pseudometric~\cite{pseudometric} that computes a distance between two cardinality vectors.

\section{Flow-Loss}
\label{sec:fl}

While P-Cost captures the impact of cardinalities on query plans, it has an important drawback as a loss function: It cannot be minimized using gradient-based methods. In fact, the gradient of P-Cost with respect to $\hatY$ is zero at almost all values of $\hatY$. To see why, notice that a small perturbation to $\hatY$ does unlikely change the path chosen by P-Opt($\hatY$); the path would only change if there were multiple cheapest paths. Therefore P-Cost will also not be affected by a small perturbation to $\hatY$. 
In this section we define an alternative to P-Opt that has a gradient w.r.t. any cardinality in the \plangraph, and use it to construct our loss function, \FlowLoss.

\subsection{From Shortest Path to Electrical Flows} 

The problem with P-Opt is that it strictly selects the shortest (cheapest) path in the plan graph. Consider, instead, the following alternative that can be thought of as a ``soft'' variant of shortest path. Assume the plan graph is an electrical circuit, with edge $e$ containing a resistor with resistance $C(e,Y)$. Now suppose we send one unit of current from $S$ to $D$. How will the current be split between the different paths from $S$ and $D$? 

In an electric circuit, paths with lower resistance\footnote{For the purpose of this discussion, we view the resistance of a path as the sum of the resistances along its edges, which corresponds to the path's {\em length} when the resistance is viewed as a distance,  or  the  path's {\em cost} when the resistance  is  viewed as the cost of an edge.} (shorter paths) carry more current, but the current does not flow exclusively on the path with least resistance. Assuming all paths have a non-zero resistance, they will all carry some current.  Importantly, every edge's  resistance affects how current is split across paths. The precise way in which current flows in the circuit can be obtained by solving the following {\em energy minimization}\footnote{Recall that the energy dissipated when current $I$ flows through a resistor with resistance $R$ is $RI^2$~\cite{circuit-book}.} problem:
\begin{align}
\text{F-Opt}(Y) = \argmin_{F} \sum_{e \in E} C(e, Y) \cdot F_e^{2} \label{eq:flows_optimization} \\
\text{s.t} \sum_{e \in Out(S)} F_e = \sum_{e \in In(D)} F_e = 1 
\label{eq:opt_constraints1} \\
\sum_{e \in Out(V)} F_e = \sum_{e \in In(V)} F_e \label{eq:opt_constraints2}
\end{align}
Here the optimization variable $F$ assigns a flow of current to each edge. 
Equation \eqref{eq:opt_constraints1} enforces that one unit of flow is sent from $S$ to $D$.
Equation \eqref{eq:opt_constraints2} is the conservation constraint for all nodes except $S$ and $D$ --- it enforces that the amount of flow going in and out of a node should be the same. The thickness of edges in Figure \ref{fig:subquery_graph} show the flows assigned to each edge by F-Opt$(\Y)$.

Computing $\text{F-Opt}$ is a classical problem in circuit design~\cite{circuit-book,maxflow1}, and it has a simple closed form expression as a function of the resistances $C(e,Y)$. We provide the exact expression and its derivation online~\cite{online_appendix}. The solution has the following form:
\begin{equation}
\text{F-Opt}(Y) = A(Y)B(Y)^{-1}i,
\numberthis
\label{eq:flows_linear_eq}
\end{equation}
where $i \in R^{N}$ is a constant vector, and $A \in R^{M,N}, B \in R^{N,N}$ are matrices with entries constructed as simple functions of the edge costs (resistances), $C(e,Y)$,  for a \plangraph with $M$ edges and $N+1$ nodes.
$\text{F-Opt}$ just multiplies two matrices, thus is clearly differentiable.
We provide an explicit closed form expression for the gradient of $\text{F-Opt}$ online~\cite{online_appendix}. The closed form expression allows us to reuse intermediate results (e.g., $B(Y)^{-1}$) to reduce the overhead of computing the gradient significantly.

We are now ready to define our final loss function.

\begin{definition}
\stitle{\FlowLoss.}
\begin{equation}
\label{eq:flow_loss}
\text{\FlowLoss} (\hatY, \Y) = \sum_{e \in E}  C(e, \Y)  \cdot %
\text{F-Opt}(e, \hatY)^{2}
\end{equation}
\end{definition}
Notice the similarity to P-Cost (Equation \ref{eq:plan_cost}). P-Cost computed the sum  of the true edge costs of the path chosen by $\text{P-Opt}(\hatY)$,  whereas \FlowLoss is a weighted sum of the true edge costs, where the weight of an edge is the square of $\text{F-Opt}(\hatY,e)$.  An alternative, intuitive interpretation of \FlowLoss is the true ``energy dissipation'' of the flows  $\text{F-Opt}(\hatY)$. This is minimized when the flows are chosen according to the true cardinalities, i.e. when $\hatY = \Y$.
Since F-Opt$(\cdot)$ and $C(\cdot)$  (Definition \ref{def:cost_model}) are both differentiable, so is \FlowLoss, and we can use the chain rule to get the gradients of \FlowLoss w.r.t $\hatY$.

\subsection{Discussion}
\label{sec:discussion}

\stitle{Beyond left-deep plans.} 
P-Cost, and by extension, \FlowLoss are defined over left-deep plans. 
Extending \FlowLoss to bushy plans is more challenging: we will need to define a graph similar to the \plangraph, where every valid bushy plan is a path, but this will lead to an exponential increase in the number of paths. 
Fortunately, it does not seem required to consider bushy plans explicitly when optimizing for cardinality estimates. First, the best left-deep plan often has reasonable performance compared to the best overall plan~\cite{qoleis}. Second, the same set of sub-plans required for optimizing a left-deep plan are also needed for bushy plans (excluding cross-joins). In particular, when indices are used, left-deep sub-plans are a prominent part of bushy plans. 
Hence, estimates that are important for choosing good left-deep plans are also important for bushy plans.

\stitle{Anchoring.} An unusual property of \FlowLoss compared to loss functions such as Q-Error is that it is not very sensitive to the {\em absolute} value of the cardinality estimates. Like an optimizer, \FlowLoss is affected more by the {\em relative} value of estimates for competing sub-plans. In particular, multiplying the cardinality estimates of all sub-plans of a query by a constant will often not change the cheapest path in the plan graph, because the costs computed using $C$ (Definition \ref{def:cost_model}) are linear in the cardinality estimates for most edges (specifically, the edges corresponding to joins that are cheaper with an index).
The implication is that training a cardinality estimation model using \FlowLoss does not ``anchor'' the learned model's outputs to the true values (e.g., it may learn to estimate cardinalities that are all roughly $5\times$ larger than the true values). It is possible to add explicit terms to the loss function that penalize large deviations from true values, or use a more precise cost model that is sensitive to absolute cardinalities.\footnote{For example, a cost model that accounts for spilling.} \FlowLoss will optimize for whichever cost model we use.
However, in our workloads we found that even without explicit anchoring, \FlowLoss learns cardinalities that perform well with PostgreSQL.

\section{Flow-Loss Analysis \label{sec:fla}}
The goal of \FlowLoss is to learn cardinality estimation models that improve query performance of a DBMS. As a concrete example, we focus specifically on improving PostgreSQL. In this section, we analyze the behavior of \FlowLoss using examples on PostgreSQL to understand how it improves on traditional loss functions like Q-Error.

\subsection{Cost Model \label{sec:fla_cm}}
P-Cost and \FlowLoss were defined using the simple cost model $C$ (Definition \ref{def:cost_model}). However, our ultimate goal is to improve query performance of PostgreSQL, which we quantify using the actual PostgreSQL cost model. 
\begin{definition}
\label{def:ppc}
\stitle{\PostgresError (\PPE).} \PPE is the same as P-Cost (Definition \ref{def:plan_cost}), but uses the default PostgreSQL cost model and exhaustive search over all plans\,---\,not only left-deep plans. To compute \PPE, we inject $\hatY$ into the PostgreSQL optimizer to get the cheapest plan (join order and physical operators) for $\hatY$. Then we cost this plan using $\Y$. We implement it using a modified version of the plugin pg\_hint\_plan\footnote{https://github.com/parimarjan/pg\_hint\_plan}~\cite{pg_hint_plan}. 
We disable materialization and parallelism in PostgreSQL as they add complexity which makes it harder to analyze.
\end{definition}

\begin{figure}[t]
    \includegraphics[width=\columnwidth]{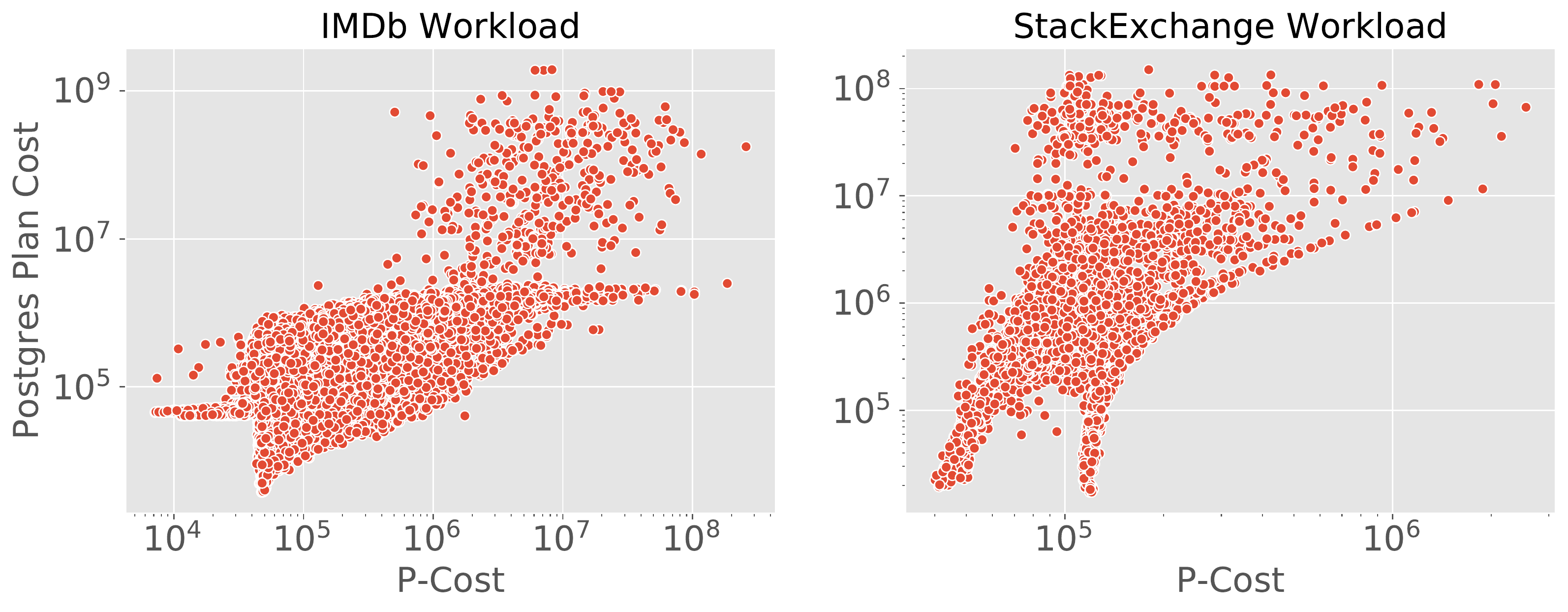}
    \caption[]%
    {{ P-Cost versus \PPE given true cardinalities for the two workloads we used. }}
    \label{fig:plancostvppc}
\end{figure}

\FlowLoss is an approximation to P-Cost, which in turn is an approximation to \PPE. For \FlowLoss to be useful, its cost model $C$ must broadly reflect the behavior of the PostgreSQL cost model.
Figure \ref{fig:plancostvppc} shows a scatter plot of P-Cost versus \PPE given true cardinalities for two workloads described in Section~\ref{sec:dataset}.
The PostgreSQL cost model includes many terms that we do not model, thus we would not expect the scale of P-Cost and \PPE to match precisely. 
Nonetheless, we observe that \PPE and P-Cost mostly follow the same trends. 
It matters less that P-Cost is not very precise, since we are merely using it as a signal to improve the cardinality estimates that lead to high costs. To optimize queries, these cardinality estimates will be provided to the PostgreSQL optimizer with its full cost model.

\subsection{Shape of Loss Functions}
\label{sec:analyze_q1}

\begin{figure}[t]
    \includegraphics[width=\columnwidth]{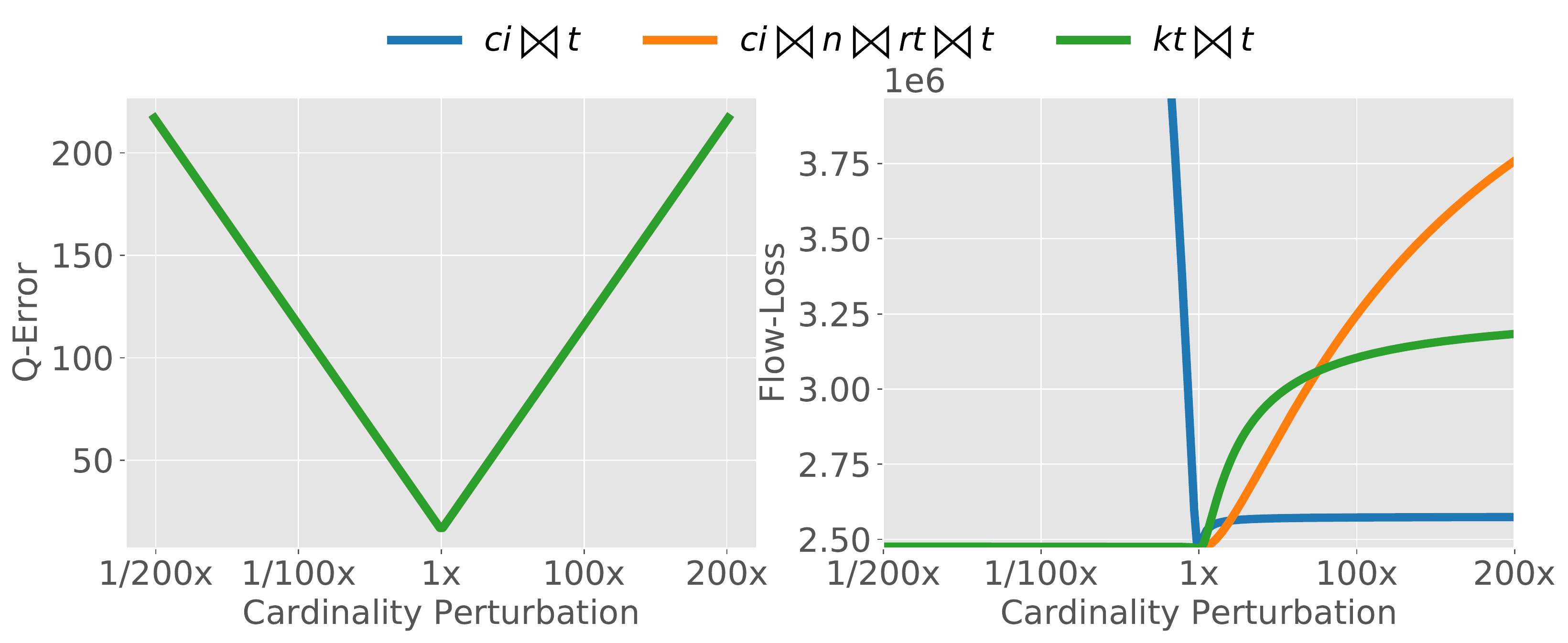}
    \caption[]
    {Comparing Q-Error (left) or \FlowLoss (right) as we vary the cardinality estimates of different sub-plans. For each data point we multiply or divide the true value (center) by $2$.}
    \label{fig:subplans_perturb}
\end{figure}

\begin{figure}[t]
    \includegraphics[width=\columnwidth]{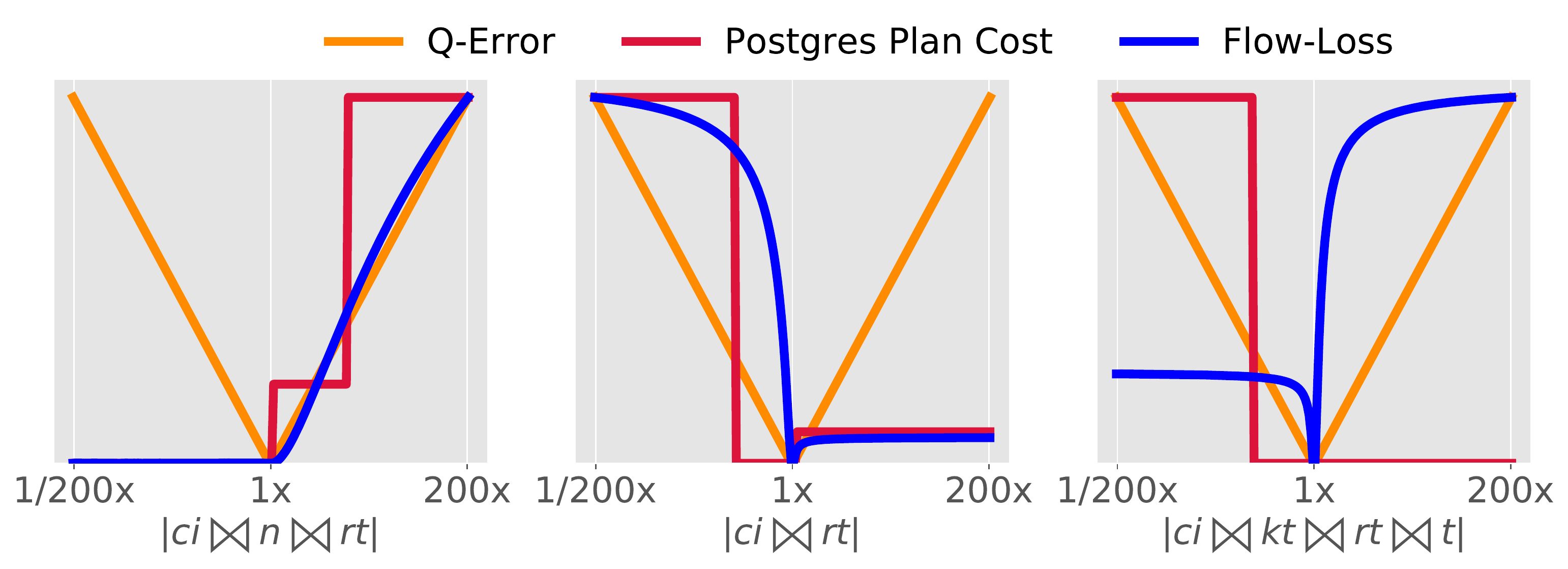}
    \caption[]
    { Comparing the shapes of Q-Error, \PPE, and \FlowLoss as we vary estimate of one sub-plan, while keeping others fixed at their true values. 
    Each loss curve is plotted with its own scale (not shown). 
    For each data point we multiply or divide the true value (center) by 2. }
    \label{fig:subquery_perturbations}
\end{figure}
\begin{figure}[ht]
  \includegraphics[width=\columnwidth]{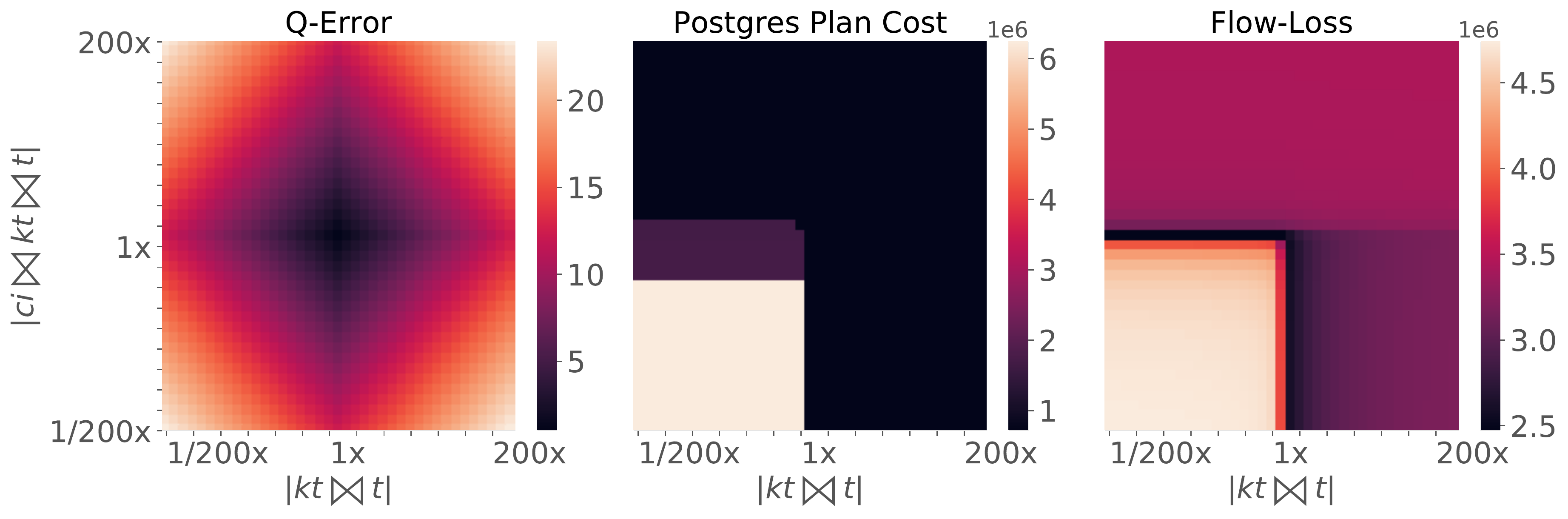}
    \caption[]
    {Comparing Q-Error, \PPE, and \FlowLoss when we vary estimates of two sub-plans at the same time. The colors go from dark (low errors) to light (high errors).}
    \label{fig:two_subquery_perturbations}
\end{figure}

Next, we will compare the behavior of Q-Error (Definition \ref{def:qerror}), \PPE (Definition \ref{def:ppc}), and \FlowLoss using our running example, query $Q_1$ (Figure~\ref{fig:q1_figure}).  
Recall that Figure \ref{fig:subquery_graph} shows the true cost of each edge, $C(e,\Y)$. As we change the cardinality of one node (sub-plan), $u$, the estimated costs of outgoing edges from $u$ will change, affecting the overall cost of any path (plan) that passes through $u$.

\stitle{\FlowLoss is sensitive to under-estimates of nodes on bad paths, and over-estimates of nodes on good paths.} Figure $\ref{fig:subplans_perturb}$ shows three representative examples of how Q-Error and \FlowLoss change as we multiply or divide the cardinality of one node by increasing amounts while keeping the others fixed at their true values.
Q-Error changes identically for all nodes (the lines overlap), but the behavior of \FlowLoss differs depending on the node. Node $ci \Join t$ has multiple expensive paths that go through it (note the red edges in Figure \ref{fig:subquery_graph}).
As we under-estimate its cardinality, \FlowLoss shoots up (blue line). This aligns with the intuition that under-estimating this node makes bad paths appear cheaper, which may cause the optimizer to choose one of them instead of the actual cheapest path.
Over-estimating its cardinality, on the other hand, make bad paths appear even more expensive, which is good as we want the optimizer to avoid these paths. Thus, it is sensible that \FlowLoss stays near its minimum in this case.
The node $ci \Join n \Join rt \Join t$ is on the cheapest path, while the node $kt \Join t$ has two relatively good paths passing through it (c.f. Figure \ref{fig:subquery_graph}).
For these nodes, \FlowLoss remains at its minimum for under-estimates (since it makes good paths appear cheaper), and shoots up for over-estimates (since it makes good paths appear more expensive). 
Recall that \FlowLoss uses all relatively good paths, not just the cheapest, and therefore, it is impacted by both nodes.

\stitle{\FlowLoss roughly tracks \PPE decision boundaries.} Figure \ref{fig:subquery_perturbations} compares the shapes of Q-Error, \PPE, and \FlowLoss as we vary the cardinality of a single node. Each curve is plotted on its own scale as we are only interested in comparing their behavioral trends. Node $ci \Join n \Join rt$ is already on the cheapest path (cf. Figure \ref{fig:subquery_graph}), so \FlowLoss is only sensitive to over-estimating its cardinality, like \PPE.
Node $ci \Join rt$ is not on the cheapest path, and like \PPE, \FlowLoss is a lot more sensitive to under-estimates as it causes flow to be diverted to the paths containing this node from potentially cheaper paths. 
Node $ci \Join kt \Join rt \Join t$ is an example of a case where \FlowLoss leads to a different behavior from PPC.
For overestimates, \PPE is flat at its minimum while \FlowLoss blows up. 
$ci \Join kt \Join rt \Join t$ is not on the cheapest path, but there are multiple nearly optimal paths using this node (cf. Figure \ref{fig:subquery_graph}). 
Since \FlowLoss routes a non-trivial amount of flow on such paths, it is sensitive to making them more expensive, even though the optimizer does not switch from the cheapest path (thus, \PPE remains flat). This is a desirable property from the standpoint of {\em robustness}. It reflects the fact that any of the nearly optimal paths could become the cheapest path and get chosen by the optimizer if the cardinalities change slightly. For instance, although node $ci \Join kt \Join rt \Join t$ is not on the cheapest path when all edges are cost using true cardinalities, it would be on the cheapest path if we underestimate the cost of the $ci \Join kt \Join t \rightarrow ci \Join kt \Join rt \Join t$ edge (or overestimate the cost of the actual cheapest path). In that case, \PPE would have been sensitive to increasing the cardinality of this node. By considering all good paths simultaneously, \FlowLoss more robustly captures the behavior of the optimizer in response to such variations in cardinalities. 
As a further example, in Figure \ref{fig:two_subquery_perturbations}, we vary cardinalities of two sub-plans simultaneously. Once again we observe that \FlowLoss roughly reflects the behavior of \PPE \,---\, it is highest when cardinalities for both the nodes are under-estimated (lower left quadrant in the figures). %

\subsection{Benefits of \FlowLoss}
\label{sec:benefits}
In practice, cardinality estimation models face several challenges: limited model capacity (making it impossible to learn all the intricacies of the data distribution), limited training data (since collecting ground truth data is expensive), insufficient features (e.g., it may be hard to represent predicates on columns with a large number of categorical values), noisy training data (cf. \S\ref{sec:wanderjoin}), changing data (e.g., Wang et al.~\cite{wang2020we} show that learned models can have a steep drop in performance after data is updated), and changing query workloads. Thus, it is inevitable that such models will make mistakes. %
As the examples in \S\ref{sec:analyze_q1} suggest, \FlowLoss guides the learning to focus on estimates that matter, and to improve their accuracy only to the extent necessary for improving query performance. This has several positive consequences as we highlight below.

\stitle{Model capacity.} Lower capacity models, or less expressive features, make it harder for learned models to achieve high accuracy. \FlowLoss helps utilizing the limited model capacity in a way that maximizes the model's impact on query performance (see ablation study in \S\ref{sec:unseen_templates}).

\stitle{Domain-specific regularization.} A model trying to minimize Q-Error treats each estimate as equally important, which makes it easy to overfit to the training data.
Regularization is a general approach to mitigate overfitting and improve generalization, but generic regularization techniques such as weight decay~\cite{wdregularization} simply bias towards learning simpler models (e.g., smoother functions) without taking advantage of the problem structure. \FlowLoss provides a stronger, guided regularization effect by utilizing domain-specific knowledge about query optimization.\footnote{There are similar examples in other ML applications, e.g., Li et al. show domain-specific loss functions for physics applications lead to improved generalization via implicit regularization~\cite{li2020kohn}.}
The key information is to know which details of the training data %
can be ignored without impacting query performance. If estimation errors on a subset of sub-plans do not typically cause worse plans, then there is no need to learn a more complex model to correct them. This is precisely what \FlowLoss does by allowing a high tolerance to cardinality estimation errors for many sub-plans (see results in  \S\ref{sec:unseen_templates}).

\stitle{Tolerance to noisy training data.} As a direct consequence of the previous point, by ignoring accuracy on less important subsets of the data, \FlowLoss can better handle noisy, or missing training data, which can let us avoid the expensive process of executing all sub-plans to generate the true cardinalities. Instead, we can train well-performing models using approximate cardinalities obtained via fast sampling techniques~\cite{wanderjoin} (c.f.  \S\ref{sec:wanderjoin}).

\section{Featurization and Models}
\label{sec:featurization}
This section introduces the model architectures that we use to evaluate \FlowLoss. 
The implementations of these models are available online\footnote{github.com/parimarjan/learned-cardinalities}.

\stitle{Featurization.} 
As described by Kipf et al.~\cite{mscn}, a sub-plan $q$ is mapped to three sets of input vectors: $T_q$, $J_q$, and $P_q$ for the tables, joins, and predicates in the sub-plan. We augment these with a vector $G_q$ that captures the properties of the sub-plan in the context of the \plangraph. A one-hot vector encodes each table in the sub-plan ($T_q$), and a second one-hot vector encodes each join ($J_q$). For \emph{{\sc range}} predicates, we use min-max normalization~\cite{mscn,deepcardinalitysurvey}. For \emph{{\sc in}} predicates we use feature hashing ~\cite{feature_hashing}. This is a standard technique in ML applications where categorical features with large alphabet sizes are hashed to $N$ bins. Even if $N$ is much smaller than the alphabet size, it still provides a signal for the learned models. For \emph{{\sc like}} predicates we use feature hashing with character n-grams~\cite{chargram}. This is useful to distinguish between extremely common and uncommon characters. For {\sc like}, we also include the number of characters and the presence of a digit as additional features. %
We find that $N=10$ bins each for every column-operator pair works well on our workloads. As proposed by Dutt et al.~\cite{lightweightmodels}, we add the cardinality estimate for each table (after applying its predicates) from PostgreSQL to that table's vector in $T_q$, which we found to be sufficient for our workload.
For a stronger runtime signal, we could add \emph{sample bitmaps}~\cite{mscn,deepsketches} (i.e., bitmaps indicating qualifying sample tuples), however, as this would significantly increase the model's parameters and hence increase memory requirements, we omit this optimization in this work.
Similarly, we do not explicitly encode {\sc group by} columns like earlier work does~\cite{learnedgroupby} and rely on PostgreSQL's estimates instead.

$G_q$ is a vector for the \plangraph-based properties of a sub-plan. 
This includes information about the immediate children of the sub-plan node in the \plangraph (i.e., the nodes obtained by joining the sub-plan with a base table). Specifically: the number of children, the PostgreSQL cost of the join producing that child, and the relative cardinality of that child compared to the sub-plan. Intuitively, such information about neighboring \plangraph nodes could be useful to generalize to new queries. We also add the PostgreSQL cardinality and cost estimate for the sub-plan to $G_q$. For all cardinalities, we apply log transformation~\cite{lightweightmodels}.

\stitle{Models.} 
To compare Q-Error and \FlowLoss, we train two representative neural network architectures with both loss functions. Fully-Connected Neural Network (FCNN) was used by Ortiz et al.~\cite{deepcardinalitysurvey} and Dutt et al.~\cite{lightweightmodels}. 
It takes as input a 1-D feature vector that concatenates the vectors in $T_q$, $J_q$, $P_q$, and $G_q$. Multi-Set Convolutional Network (MSCN) was proposed by Kipf et al.~\cite{mscn} based on the DeepSets architecture~\cite{deepsets}, and we extend it to include the $G_q$ features as well. These are very different architectures, and represent important trade-offs --- FCNN is a lightweight model that trains efficiently, but does not scale to increasing database sizes (number of neural network weights grow with the number of columns), while MSCN uses a set-based formulation that is scalable but is less efficient to train.

\section{\OurDataset (\CEB)}
\label{sec:dataset}
\subsection{Dataset}

\begin{table}[tbh]
    \small
    \centering
    \caption{Comparing \CEB with JOB.}
    \renewcommand{\arraystretch}{1.2}
    \begin{tabular}{@{}llll@{}}
    \toprule
    \textbf{Dataset} 
    & \parbox{1cm}{\textbf{   JOB \\(IMDb)}}
    & \parbox{1cm}{\textbf{ \CEB \\(IMDb)}}
    & \parbox{1cm}{\textbf{\CEB \\(SE)}} \\
    \hline
        \# Queries & 113 & 13,644 & 3435 \\
        \# Sub-plans & 70K & 3.5M & 500K \\
        \# Templates & 31 & 15 & 6 \\
        \# Joins  & 5 -- 16 & 5 -- 15 & 5 -- 8 \\
        \# Optimal plans & 88 & 2200 & 113 \\
    \bottomrule
    \end{tabular}
    \label{tab:dataset}
\end{table}

\begin{figure}[ht]
\centering
\includegraphics[width=0.9\columnwidth]{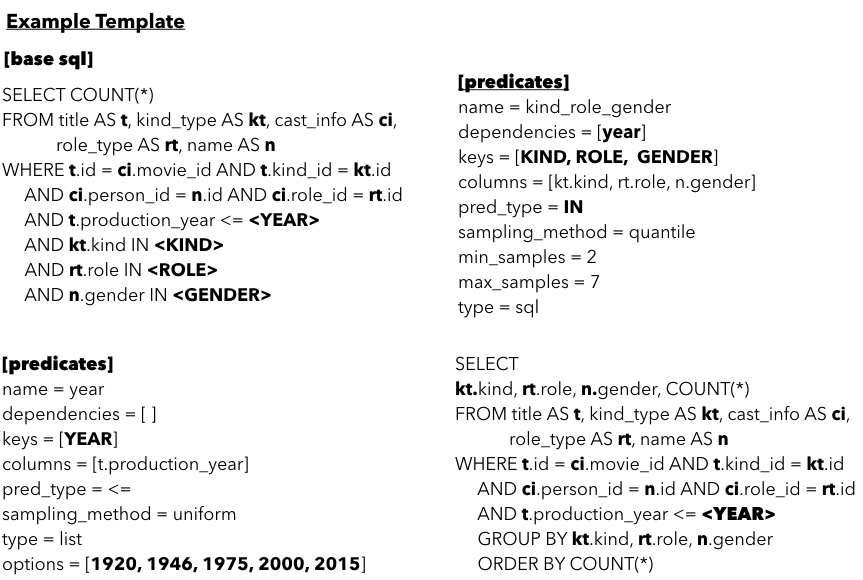}
\caption{TOML configuration file for generating queries based on a predefined template and rules. }
\label{code:toml_template1}
\end{figure}

\stitle{Benchmark.} We create a tool to generate a large number of challenging queries based on predefined templates and rules.
Using this tool, we generate the \OurDataset (\CEB)~\cite{our_dataset}, a workload on two different databases (IMDb~\cite{qoleis} and StackExchange (SE)~\cite{stackexchange}) containing over $16K$ unique queries and true cardinalities for over $4M$ sub-plans including {\sc count} and {\sc group by} aggregates, and {\sc range}, {\sc in}, and {\sc like} predicates. Table \ref{tab:dataset} summarizes the key properties of \CEB, and contrasts them with Join Order Benchmark (JOB)~\cite{qoleis}. Notice that for the $13K$ IMDb queries in \CEB, there are over $2K$ unique plans generated by PostgreSQL with true cardinalities --- showing that different predicates lead to a diverse collection of optimal query plans.
\CEB addresses the two major limitations of queries used in previous works~\cite{mscn,deepcardinalitysurvey,lightweightmodels2}: First, past work on supervised cardinality estimation~\cite{mscn,lightweightmodels2,deepcardinalitysurvey} evaluate on  workloads with only up to six joins per query. \CEB has much more complex queries ranging from five to sixteen joins. Second, while JOB~\cite{qoleis} contains challenging queries with up to $16$ joins, they only have two to five queries per template. This is insufficient training data for supervised learning methods. \CEB contains hundreds of queries per hand-crafted template with real-world interpretations.

\stitle{Query generator.} Generating predicate values for query templates is challenging because predicates interact in complex ways, and sampling them independently would often lead to queries with zero or very few results.
Our key insight is to generate interesting predicate values for one, or multiple columns together, using predefined SQL queries that take into account correlations and other user specified conditions.
Figure \ref{code:toml_template1} shows a complete template which generates queries with the same structure as our running example, $Q_1$. 
We will walk through the process of generating a sample query following the rules specified in this template.
[base sql] is the SQL query to be generated, with a few unspecified predicates to be filled in.
[predicates] are rules to choose the predicates for groups of one or more columns. The predicate {\sc year} is of type less than or equal to, and we choose a value uniformly from the given list. 
We sample filter values for the remaining three {\sc in} predicates together because {\sc kind}, {\sc role}, and {\sc gender} are highly correlated columns. For these, we also add {\sc year} as a dependency --- as the year chosen would influence predicate selectivities for all these columns. We generate a list of candidate triples 
using a {\sc group by} query. 
From this list, we sample $2$ to $7$ values for each {\sc in} predicate.

\stitle{Timeouts.} Some sub-plans in the StackExchange queries time out when collecting the true values. This is due to unusual join graphs which make certain sub-plans behave like cross-joins (see online appendix~\cite{online_appendix}). In such cases, we use a large constant value in place of the true cardinalities as the label for the timed out sub-plans in the training data. We verified that the plans generated by injecting all known true cardinalities and this constant value into PostgreSQL leads to almost $10\times$ faster runtimes than using the default PostgreSQL estimates. Thus, despite the timeouts, our labels for StackExchange are a good target for training a cardinality estimation model. %

\subsection{Approximate Training Data}
\label{sec:wanderjoin}

Intuitively, we may not need precise cardinality estimates to get the best plans --- thus, approximate query processing techniques, such as \emph{wander join}~\cite{wanderjoin} or IBJS~\cite{ibjs}, should provide sufficient accuracy. However, we cannot use these techniques for query optimization because they are too slow to provide estimates for all sub-plans at runtime. But these techniques are much faster than generating the ground truth cardinality estimates for all sub-plans, which is by far the most expensive step in building a cardinality estimation model.
We hence propose a new version of the wander join algorithm that is suited for the task of efficiently generating labeled training data for a large number of similar queries.

Wander join estimates the cardinality of a given query by doing random walks on the join graph: you start with a random tuple in the first table, and look up a tuple with a matching join key in the second table using an index, and so on. 
The walks may fail either due to predicates, or no matching join keys.
Our key modification is to first apply the predicates on the base tables, and then do the random walks. This does not change the theory around wander join, and makes the walks significantly faster. But, the cost of applying the predicates first is too expensive for AQP applications --- to efficiently do the random walks, one would also need to create an index on the tables after applying the predicate. But, for generating the training data, it is a one-time cost --- and then, we can reuse the filtered tables for all sub-plans in the query. Moreover, because \CEB consists of predicates that are often re-used in different combinations across queries, we are able to re-use the filtered tables for many queries with the same template.

We use this only as a proof of concept; our implementation is not optimized, and uses a mix of Python and SQL calls to do the random walks. Despite this, we generate the wander join estimates with speedups over generating ground truth data that range from $10\times$ to $100\times$ for different templates. For instance, for the largest template with around $3K$ sub-plans, generating all the ground truth data on a single core takes about \emph{$5$ hours}, while wander join estimates take less than \emph{$5$ minutes} on average, and give almost equally good plans. In Section~\ref{sec:seentemplates}, we explore if the wander join estimates are as good as true cardinalities to train learned models.\footnote{This learning problem is similar to how obfuscation by adding random noise to training data is used to learn ML models while preserving privacy~\cite{zhang2018privacy}.}

\section{Experiments}

\begin{figure*}[t]
    \begin{subfigure}[b]{\columnwidth}
        \centering
        \includegraphics[width=\textwidth]{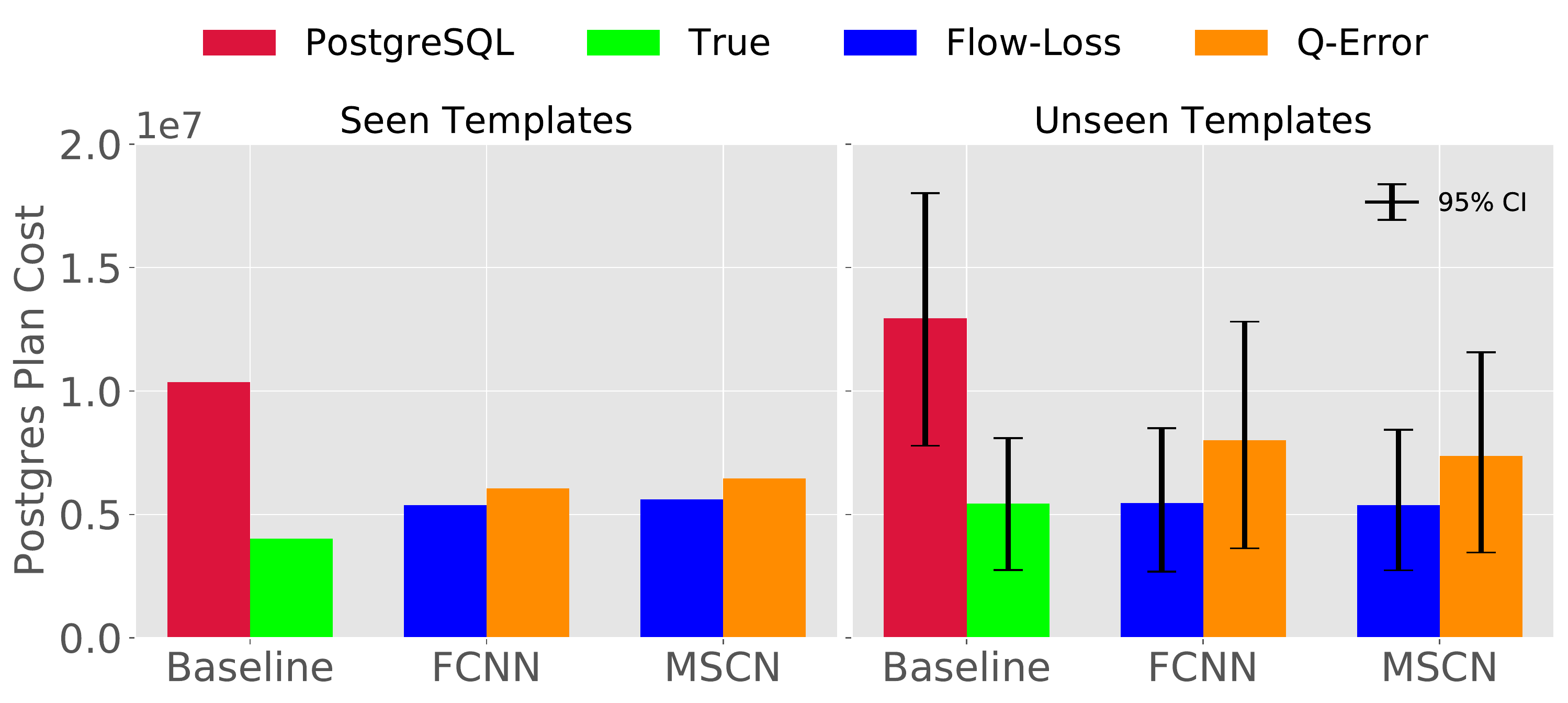}
        \caption[]%
        {{\small Mean \PostgresError (\PPE).}}
        \label{fig:all_exp_summary_plan}
    \end{subfigure}
    \hfill
    \begin{subfigure}[b]{\columnwidth}  
        \centering 
        \includegraphics[width=\textwidth]{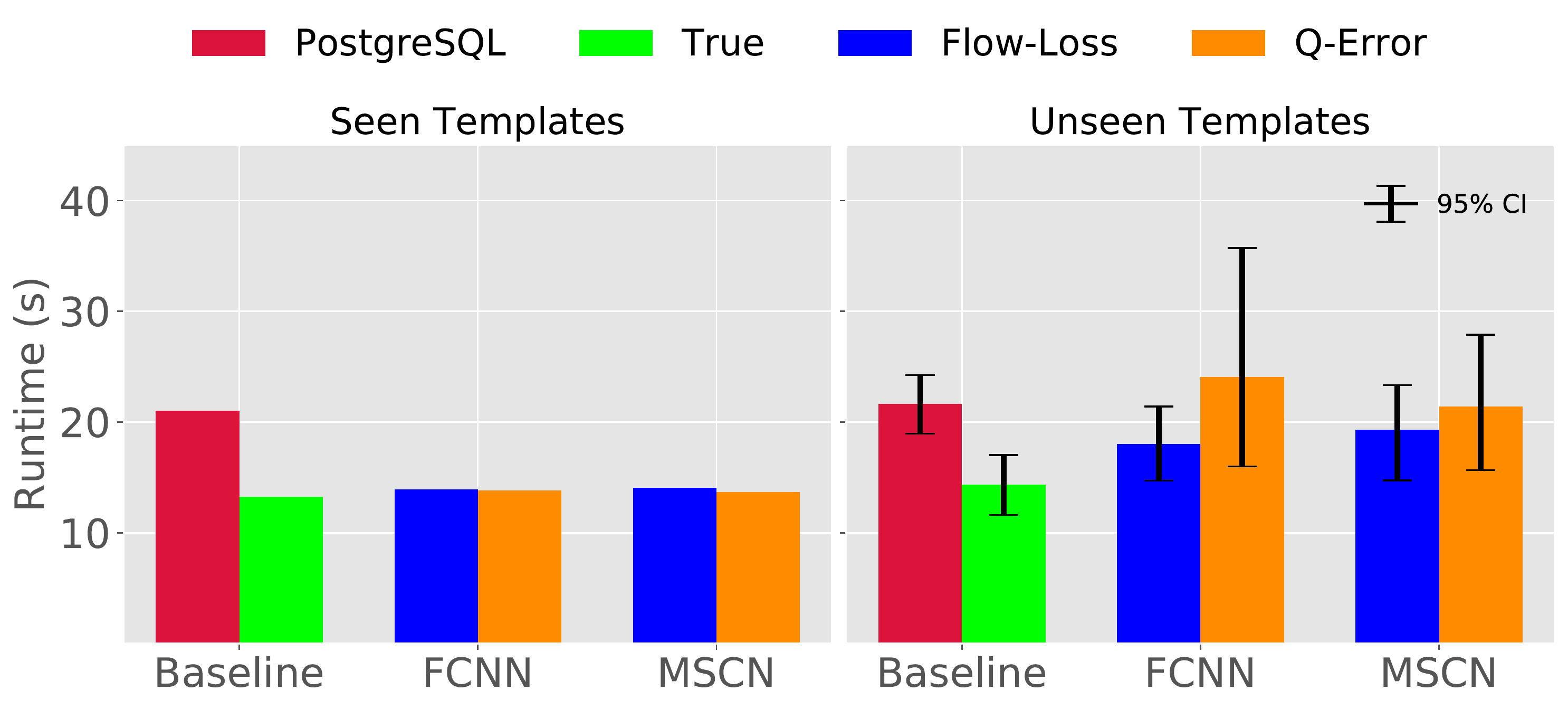}
        \caption[]%
        {{\small Mean query runtimes.}}
        \label{fig:all_exp_summary_runtime}
    \end{subfigure}
    \caption[]%
    {{ Comparing performance of all models on seen versus unseen templates. For unseen templates, we do ten experiments using ten different training/test template splits.
    }}
    \label{fig:all_exp_summary}
\end{figure*}

\stitle{Setup.} We use PostgreSQL 12 with parallelism and materialization disabled for all our experiments. For the runtime experiments, we use an Amazon EC2 instance with a NVMe SSD hard drive, and 8GB RAM (m5ad.large for IMDb, and m5ad.xlarge for StackExchange).

\stitle{Loss functions.} Our main focus is to compare the Q-Error and \FlowLoss loss functions to train the neural network models. We also compare with \Priority ~\cite{costguided}, which was our exploratory earlier work in tweaking the Q-Error loss function to focus on queries which have high \PPE. We use the true cardinalities and estimates from PostgreSQL as baselines to compare the learned  models.

\stitle{Training and test sets.} We consider two scenarios:
\begin{enumerate}[noitemsep,topsep=0pt,leftmargin=5mm,itemsep=0mm,itemindent=0mm]
    \item \textbf{Testing on seen templates.} The model is evaluated on new queries from the same templates that it was trained on.
    We put $20\%$ of the queries of each template into the validation set, and $40\%$ each into the training and test sets. The hyperparameters are tuned on the validation set, and we report results from the test set.
    \item \textbf{Testing on unseen templates.} The model is evaluated on different templates than the ones it was trained on. We split the templates equally into training and test templates. Since the number of templates is much smaller than the number of queries, we use ten-fold cross-validation for these experiments: the training / test set splits are done randomly using ten different seeds (seeds = $1-10$). We use the same hyperparameters as determined in the seen templates scenario. Even though the templates are different in the second scenario, there would be a significant overlap with the training set on query \emph{sub-plans}. This tests the robustness of these models to slight shifts in the workload.
\end{enumerate}

\stitle{Key results.} Figure \ref{fig:all_exp_summary_plan} shows the results of all approaches w.r.t. \PPE on IMDb. All models outperform PostgreSQL's estimator significantly on seen templates. However, only the \FlowLoss trained models do so consistently on unseen templates as well. For seen templates, the models trained using \FlowLoss do better than the models trained using Q-Error on \PPE. All models get worse when evaluated on unseen templates - but the \FlowLoss models degrade more gracefully. When the queries are from seen templates, the difference in \PPE does not translate into runtime improvements (cf. Figure~\ref{fig:all_exp_summary_runtime}). However, on unseen templates, we see clear improvements in runtime as well. %

\subsection{Testing on seen templates}
\label{sec:seentemplates}

\begin{table*}[t]
\small
\centering
\caption{Models trained and evaluated on the same IMDb templates. We show $\pm 1stddev$ for Q-Error and \PPE from three repeated experiments, and execute plans from one random run. $Xp$ refers to the $Xth$ percentile.}
\setlength{\tabcolsep}{3pt}
\ra{1.2}
\begin{tabular}{@{}rccc ccc ccc@{}}
\toprule
\\&
\multicolumn{3}{c}{\hspace{-5ex} \textbf{Q-Error}} & 
\multicolumn{3}{c}{\hspace{-3ex}\textbf{Postgres Plan Cost (Millions)}} & 
\multicolumn{3}{c}{\textbf{Runtime}} \\
\cmidrule(r{0.6cm}){2-4}
\cmidrule(r{0.3cm}){5-7}
\cmidrule(r{0.0cm}){8-10}

 & \centering 50p & 90p & \centering 99p & Mean & 90p & 99p & Mean & 90p & 99p  \\  
\midrule

\textbf{Baselines} & \\
True &   1 & 1 &     1  & \textbf{4.7} & 1.1  & 131.5 &   \textbf{13.23} &  28.96 &   60.83 \\ 
PostgreSQL &  42.87 &   48K &   2.2M  & 10.7 & 9.6 & 269.7 &   20.86 &   46.86 &  101.57  \\
\midrule 

\textbf{FCNN} & \\
Q-Error & \textbf{2.0 $\pm$ 0.1} & 10.5 $\pm$ 1.6 & 0.1K $\pm$ 32.0 & 6.4 $\pm$ 0.5 & 1.8 $\pm$ 0.1 & 142.6 $\pm$ 11.7 & \textbf{13.83} &  30.82 &  60.10  \\ 

\Priority 
& 2.5 $\pm$ 0.1 & 17.1 $\pm$ 1.9 & 0.4K $\pm$ 83.9 &
\textbf{5.9 $\pm$ 0.4} & 1.7 $\pm$ 0.0 & 135.8 $\pm$ 18.9 & 
13.91 & 31.48 &  61.40  \\

\FlowLoss &  4.0 $\pm$ 0.9 & 83.4 $\pm$ 36.3 & 4.1K $\pm$ 1.9K & 
6.0 $\pm$ 0.7 & 1.8 $\pm$ 0.1 & 136.0 $\pm$ 23.9 & 
13.93 & 30.97 &  60.50   \\ 

\midrule 

\textbf{MSCN} & \\
 Q-Error &  \textbf{2.0 $\pm$ 0.0} & 9.6 $\pm$ 0.2 & 0.1K $\pm$ 3.9 & 
 6.4 $\pm$ 0.3 & 1.9 $\pm$ 0.1 & 189.1 $\pm$ 6.6 & 
 \textbf{13.69} & 30.49 &  58.49  \\ 
\Priority & 2.8 $\pm$ 0.4 & 25.2 $\pm$ 8.3 & 0.7K $\pm$ 0.4K & 
6.1 $\pm$ 0.4 & 2.4 $\pm$ 0.4 & 163.8 $\pm$ 10.1 & 
13.83 & 31.10 &  60.00 \\
\FlowLoss & 3.0 $\pm$ 0.1 & 44.4 $\pm$ 5.6 & 2.1K $\pm$ 0.4K & 
\textbf{5.5 $\pm$ 0.6} & 2.2 $\pm$ 0.1 & 161.9 $\pm$ 16.2 & 
14.08 & 31.61 &  59.88    \\

\bottomrule
\end{tabular}
\label{tab:imdb-all-test}
\end{table*}

Table \ref{tab:imdb-all-test} summarizes the results for the IMDb workload when trained and tested on all IMDb templates. Each experiment is repeated three times, and we show $\pm 1stddev$ for each statistic. All learned models improve significantly over PostgreSQL on all metrics, and do about equally well.

\stitle{Worse Q-Error, better \PPE, similar runtimes.} On Q-Error, the models trained using \FlowLoss do worse than the models trained using Q-Error. This is to be expected --- our goal was to improve cardinality estimates only when its crucial for query planning. 
The \FlowLoss trained models distinctly improve mean \PPE over the Q-Error models, getting close to the \PPE with true cardinalities. This suggests that \FlowLoss models better utilize their model capacity to focus on sub-plans that are more crucial for \PPE. It also shows that better Q-Error estimates do not directly translate into improved plans. However, in terms of runtimes, all models do equally well, and are very close to the optimal.

\begin{figure}[t]
    \centering
    \includegraphics[width=0.54\columnwidth]{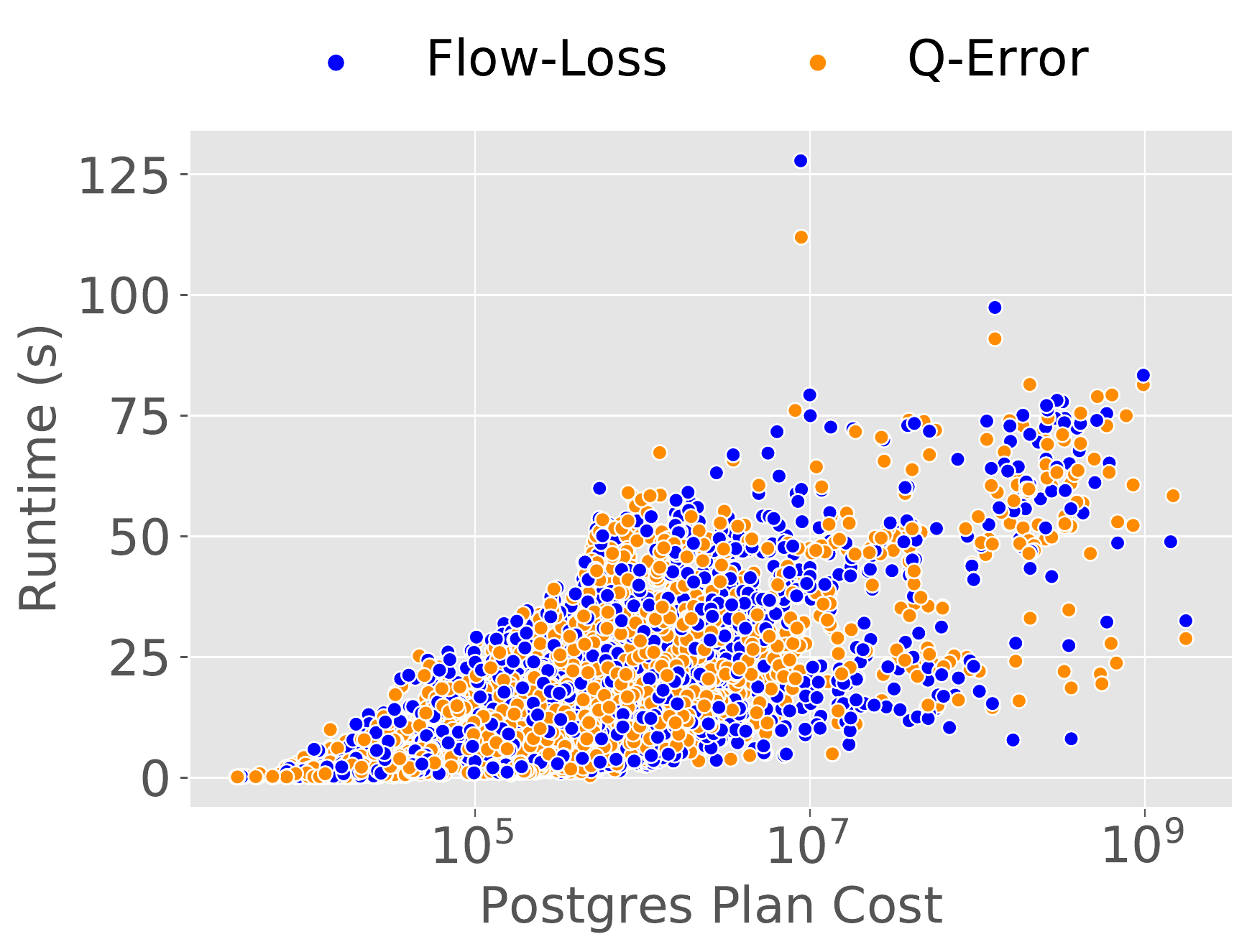}
    \caption[]%
    {{ \PPE versus runtimes for MSCN models trained with Q-Error or \FlowLoss and evaluated on seen templates. }}
    \label{fig:costvruntime}
\end{figure}
\stitle{\PPE versus runtimes.}
Figure \ref{fig:costvruntime} shows the trends for \PPE and runtimes are roughly correlated. Notice that the costs (x-axis) are shown on a log scale --- thus, \emph{order of magnitude better costs translate to faster runtimes.}

\begin{figure}[t]
\includegraphics[width=\columnwidth]{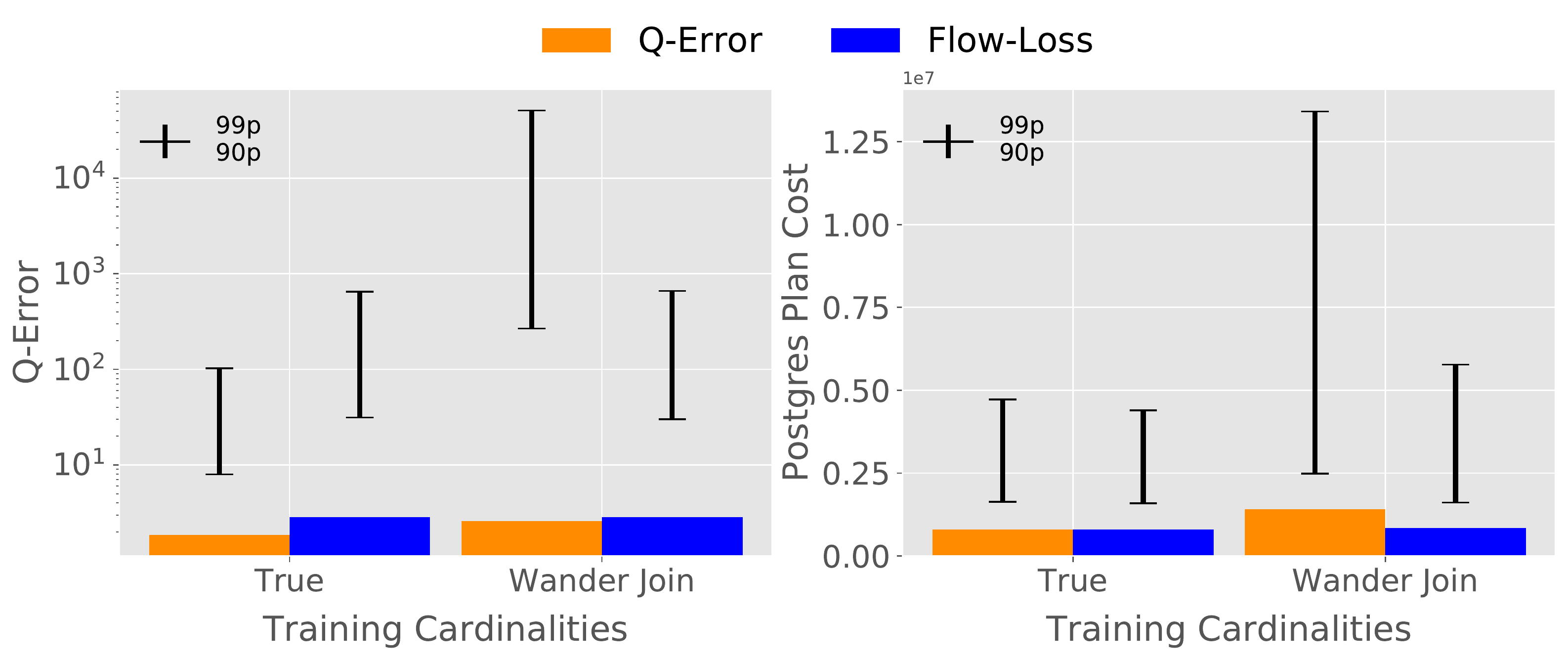}
\caption{Median Q-Error and mean \PPE 
for FCNN model trained with true or wander join cardinalities.}
\label{fig:wanderjoin}
\end{figure}
\stitle{Training with wander join estimates.}
Figure \ref{fig:wanderjoin} shows the Q-Error and \PPE for models trained with true cardinalities and with wander join estimates (cf. \S~\ref{sec:wanderjoin}) for a subset of nine templates (that do not include {\sc like} predicates) from the \CEB workload. Notice that training with wander join estimates is almost as good as training with true cardinalities. And, \FlowLoss models are robust when trained using the noisy wander join estimates --- meanwhile, the Q-Error models trained using wander join estimates have a much worse tail performance on both estimation accuracy (Q-Error) and \PPE.
This supports our hypothesis that the models trained using \FlowLoss are able to avoid overfitting to noisy data that may not be as relevant for query optimization (cf. \S\ref{sec:benefits}).

\subsection{Testing on unseen templates}
\label{sec:unseen_templates}

\begin{figure}[t]
    \begin{subfigure}[b]{\columnwidth}
        \includegraphics[width=\textwidth]{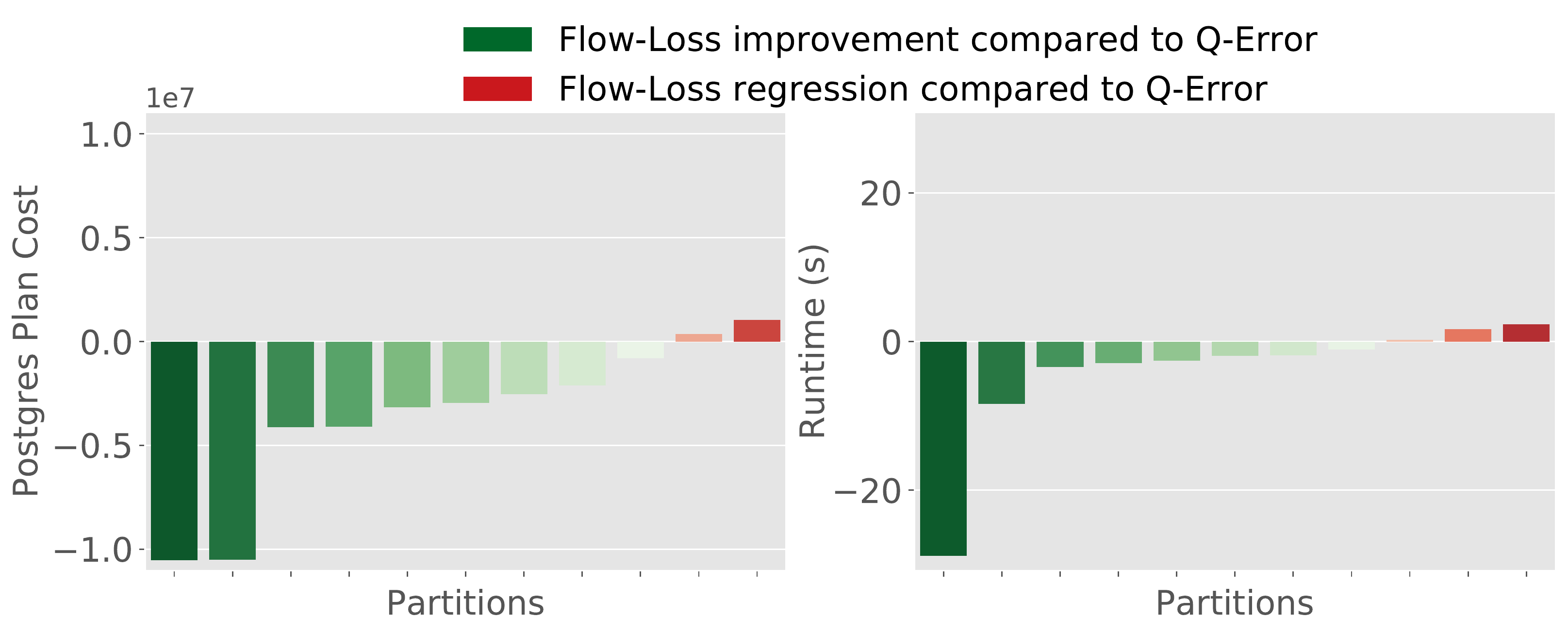}
        \caption[]%
        {{IMDb workload, FCNN or MSCN model.}}    
        \label{fig:meandiffs1}
    \end{subfigure}
    \begin{subfigure}[b]{\columnwidth}
        \includegraphics[width=\textwidth]{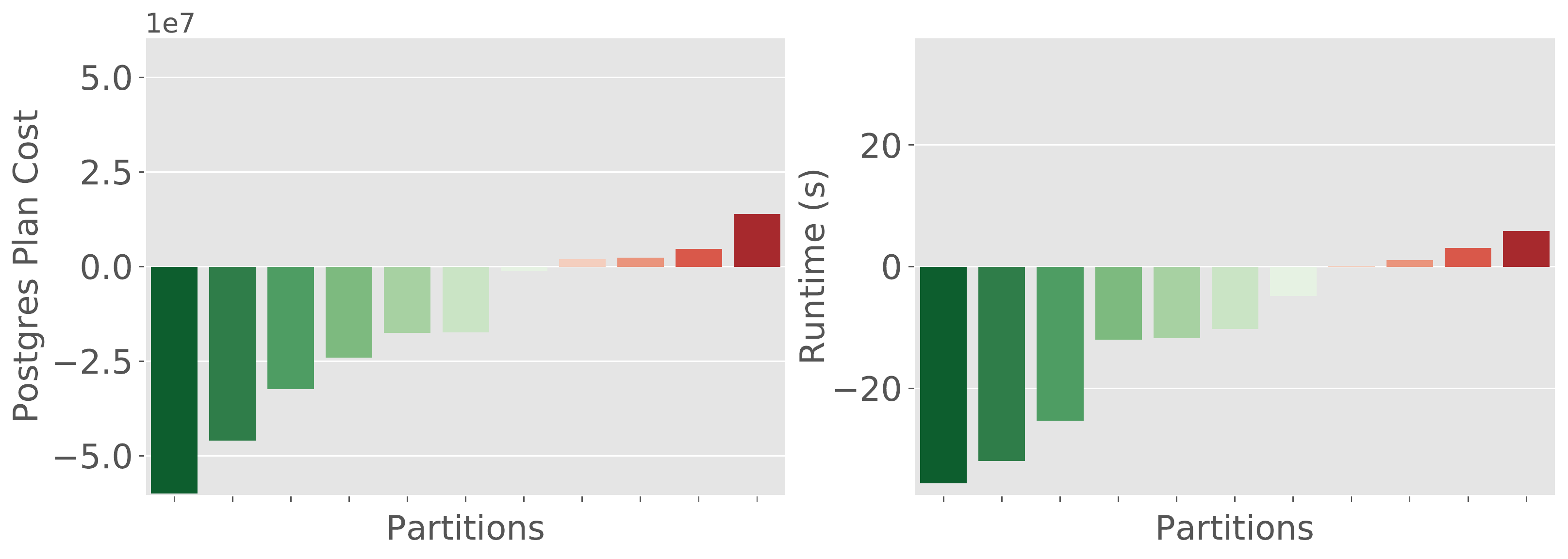}
        \caption[]%
        {{ StackExchange workload, MSCN model.}}   
        \label{fig:meandiffs2}
    \end{subfigure}
    \caption[]%
    {{ 
    Each bar shows the \PPE or runtime improvement/regression of \FlowLoss over Q-Error on an unseen partition and the same model.
    \emph{Lower is better.}
    }}
\end{figure}

\begin{figure}[t]
    \includegraphics[width=\columnwidth]{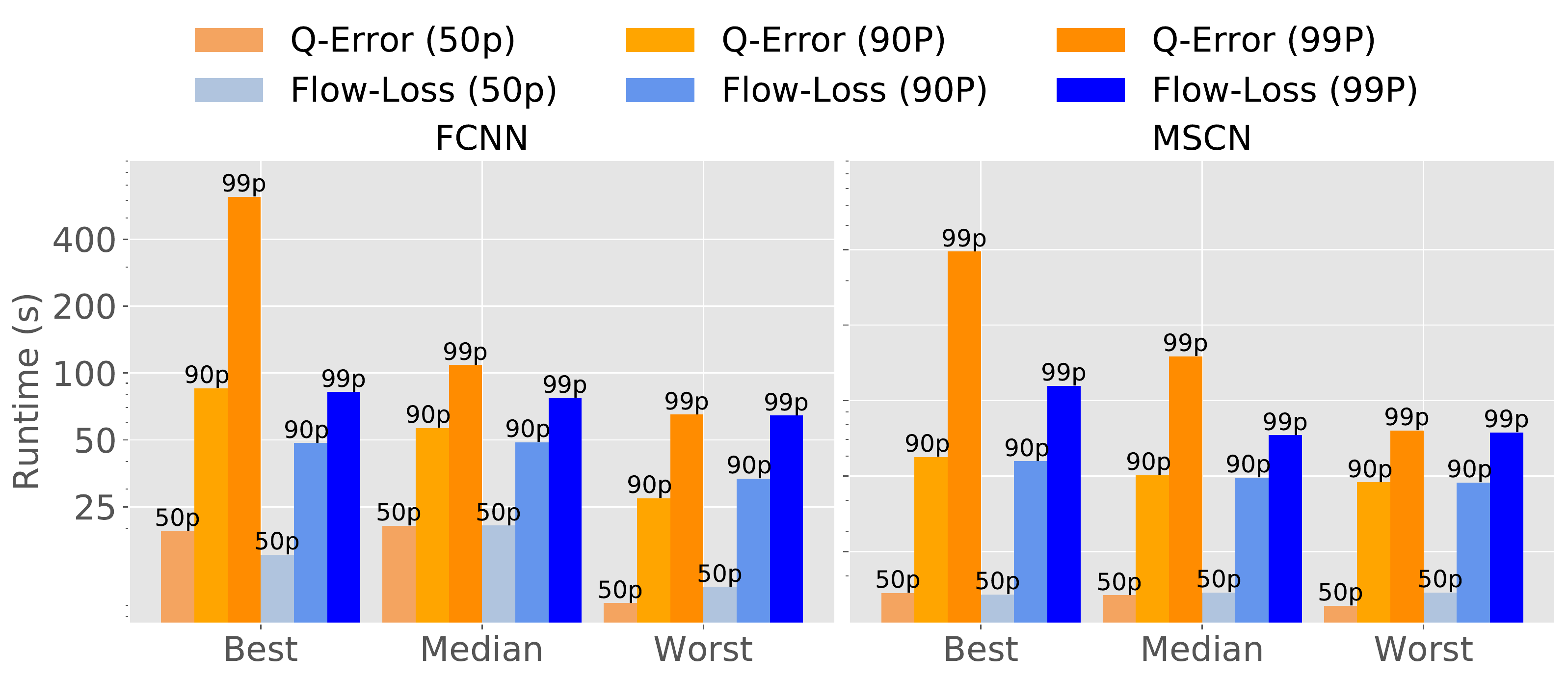}
    \caption[]%
    {{ Best, median, and worst partition for \FlowLoss models by runtime difference from its corresponding Q-Error model on unseen template partitions.
    }}
   \label{fig:exp_diff_rts}
\end{figure}

When we split the training set and test set by templates, each partition leads to very different information available to the models --- therefore we will analyze the partitions individually.
We only show results for the models trained with \FlowLoss or Q-Error (\Priority does clearly worse than \FlowLoss w.r.t. \PPE, see online appendix~\cite{online_appendix}).

\stitle{\FlowLoss generalizes better.} In Figure \ref{fig:meandiffs1}, we look at the performance of a model trained with \FlowLoss compared to one trained with Q-Error w.r.t. \PPE and query runtime. A single bar represents the same model architecture (FCNN or MSCN) trained and evaluated on one of the ten partitions in the unseen templates scenario. To avoid the runtime overhead, we omit the experiments where the \PPE of the models trained with Q-Error and \FlowLoss were within $500K$ of each other.
This figure highlights the overall trends across all unseen partition experiments: we see significant improvements on some partitions, relatively smaller regressions, and similar performance on many partitions.

\stitle{Zooming in on partitions.} For the FCNN and MSCN models, we sort all the partitions by the difference in the mean runtimes between the \FlowLoss and the Q-Error models. We select the best, median, and worst partition for \FlowLoss and show the $50p$, $90p$, and $99p$ for runtimes in Figure \ref{fig:exp_diff_rts}. For both architectures, the model trained with \FlowLoss significantly improves on all percentiles for the best partition, and has about the same performance on the worst partition. Even on the median partition, the \FlowLoss model does clearly better at the tail.

\begin{figure}[t]
    \centering
        \includegraphics[width=\columnwidth]{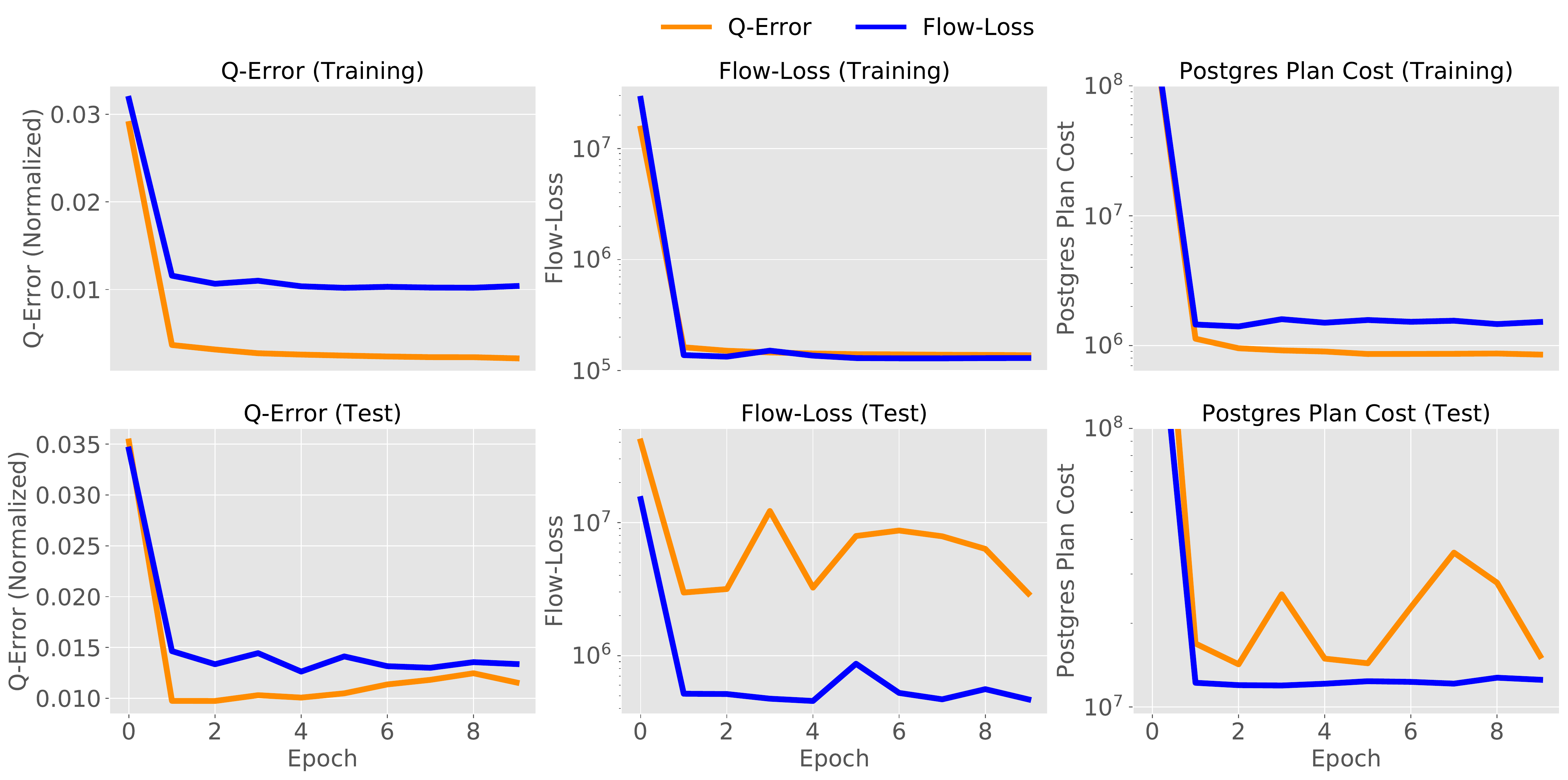}
        \caption[]%
        {{ Learning curves for one unseen templates partition
        showing mean of all metrics for MSCN models.
        }}
    \label{fig:learning_curves}
\end{figure}
\stitle{Learning curves.} Figure \ref{fig:learning_curves} shows the MSCN model's learning curves for Q-Error (normalized while training as done by Dutt et al.~\cite{lightweightmodels}), \FlowLoss, and \PPE on one partition (seed = 7) trained using Q-Error or \FlowLoss.
We see that the Q-Error model has smooth training set curves for all metrics, but it behaves erratically on the test set. This is because it is trying to minimize estimation accuracy, but since the test set contains queries from unseen templates, it is much more challenging.
Note that the \FlowLoss curves closely resemble the \PPE curves.
This similarity is particularly obvious for the Q-Error model on the test set.
In \S \ref{sec:analyze_q1}, we showed simple examples where the \FlowLoss metric closely tracked the \PPE. This shows that \FlowLoss can track the \PPE well even in more complex scenarios.
Notice that on the test set, in terms of Q-Error, the \FlowLoss model is consistently worse than the Q-Error model, while it is much better on the \FlowLoss and \PPE metrics. This suggests the \FlowLoss model is using its capacity better to focus on \PPE instead of minimizing estimation accuracy.

\begin{figure}[t]
    \centering
        \includegraphics[width=\columnwidth]{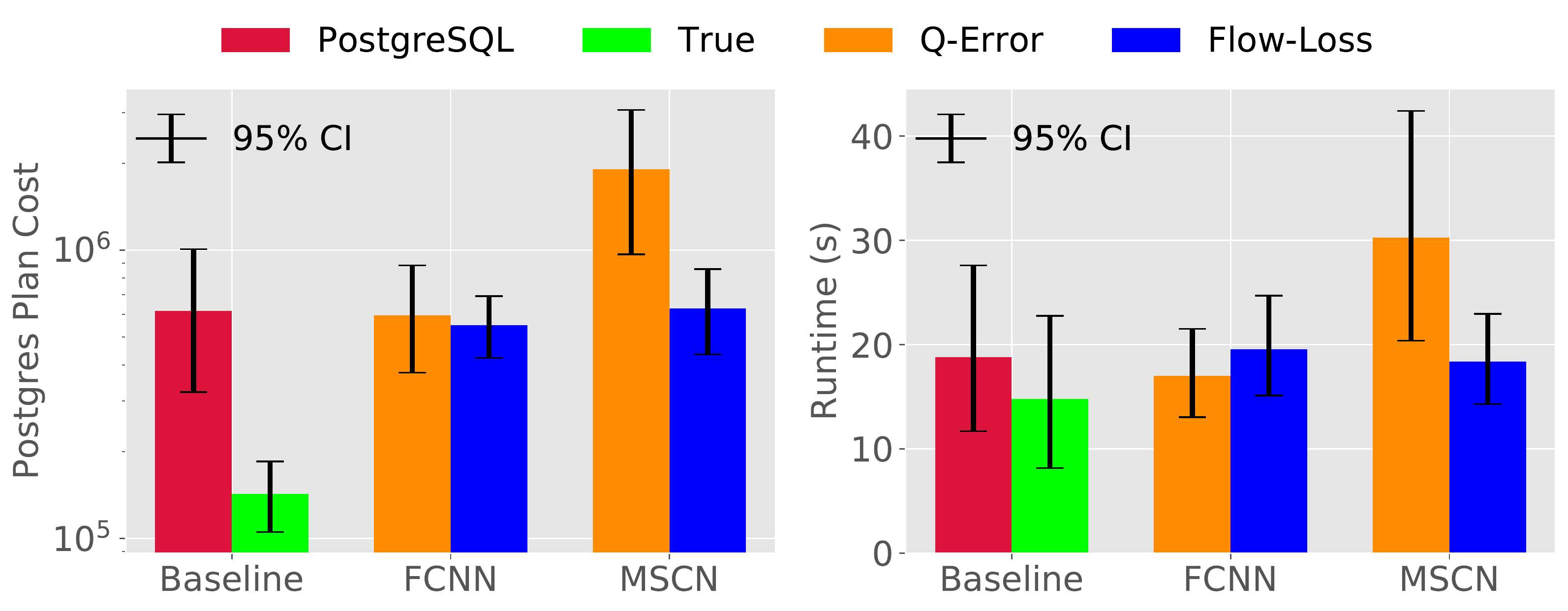}
        \caption[]%
        {{ Mean \PPE and runtimes for all models trained with Q-Error or \FlowLoss on \CEB, and evaluated on JOB.
        }}
    \label{fig:job_summary}
\end{figure}

\stitle{Join Order Benchmark.} JOB is not suitable for training a supervised learning model as it has too few queries. But, we can use it as an evaluation set for a model trained on all the templates from \CEB. This is similar to the unseen templates scenario: The JOB queries are less challenging in terms of \PPE (for instance, PostgreSQL estimates have $20\times$ lower mean \PPE on JOB than \CEB). However, they are more diverse: JOB has $31$ templates, and includes predicates on columns not seen in \CEB. We train on queries from all the \CEB templates and evaluate on the JOB queries.
Figure \ref{fig:job_summary} summarizes the results of the \FlowLoss and Q-Error models for both architectures over three repeated runs.
Both \FlowLoss models improve slightly on \PPE over PostgreSQL while achieving similar runtimes.
The FCNN model trained with Q-Error performs similarly, but the MSCN model trained with Q-Error shows much higher variance and does significantly worse. We use this experiment as a sanity check to show that even when the queries are very different, our models avoid disastrously bad estimates.

\begin{figure}[t]
    \centering
    \includegraphics[width=0.54\columnwidth]{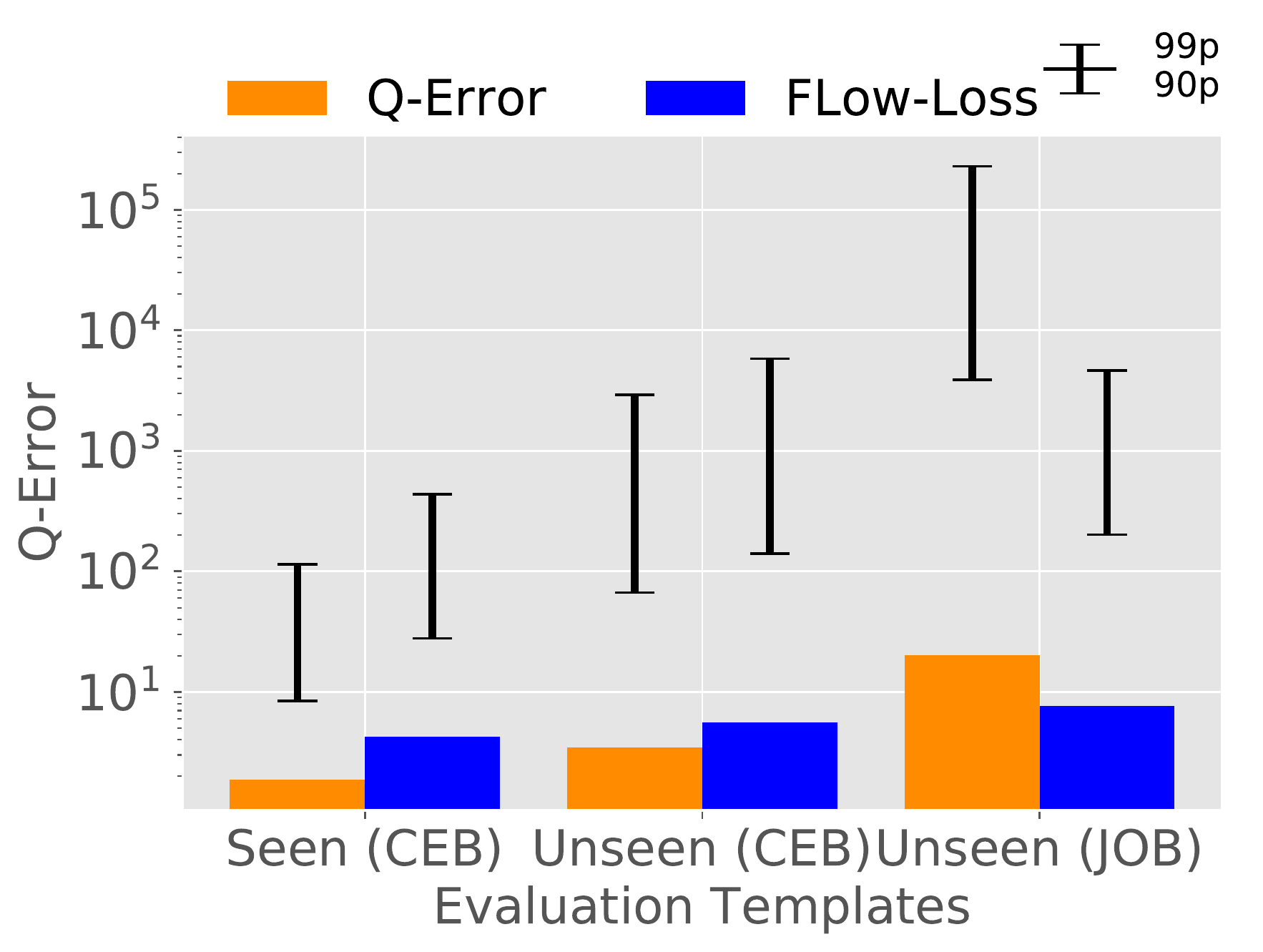}
    
    \caption[]%
    {{ Median, 90p, and 99p Q-Error for FCNN models trained on queries from \CEB and evaluated on seen templates, unseen templates, and JOB.
    }}
    \label{fig:qerr_seenvsunseen}
\end{figure}

\stitle{Domain specific regularization effect.} Figure \ref{fig:qerr_seenvsunseen} shows the median, $90p$, and $99p$ Q-Errors for the three scenarios we have looked at.
We show results for the FCNN architecture and omit MSCN here, which performs very similar.
For seen templates, the models trained with Q-Error only go up to $100$ at the $99p$ of millions of unseen sub-plans. To achieve such low estimation errors, the model needs to get quite complex, and overfit to noisy patterns, like precise estimates for {\sc ilike} predicates or for sub-plans with $10$ tables that may anyway get pruned during the dynamic programming optimizer search. Often, we do not need such precise estimates.
Models trained with \FlowLoss achieve better \PPE despite an order of magnitude higher Q-Errors at the $99p$, which suggests that it learns a simpler model that seems more effective for the task of query optimization. More strikingly, as we consider the unseen templates scenario --- the models trained with Q-Error get almost $10\times$ worse at the $90p$ and $99p$, while the models trained with \FlowLoss only get about $1.5\times$ worse. This pattern continues on to the JOB templates --- where the \FlowLoss models even have better estimation accuracy than the Q-Error models.
This supports our regularization hypothesis (cf. \S \ref{sec:benefits}), and shows that the \FlowLoss models can avoid overfitting in a way that does not harm its performance on \PPE, but the simpler models make it generalize better to changing query patterns.

\begin{figure}[t]
\centering
\includegraphics[width=\columnwidth]{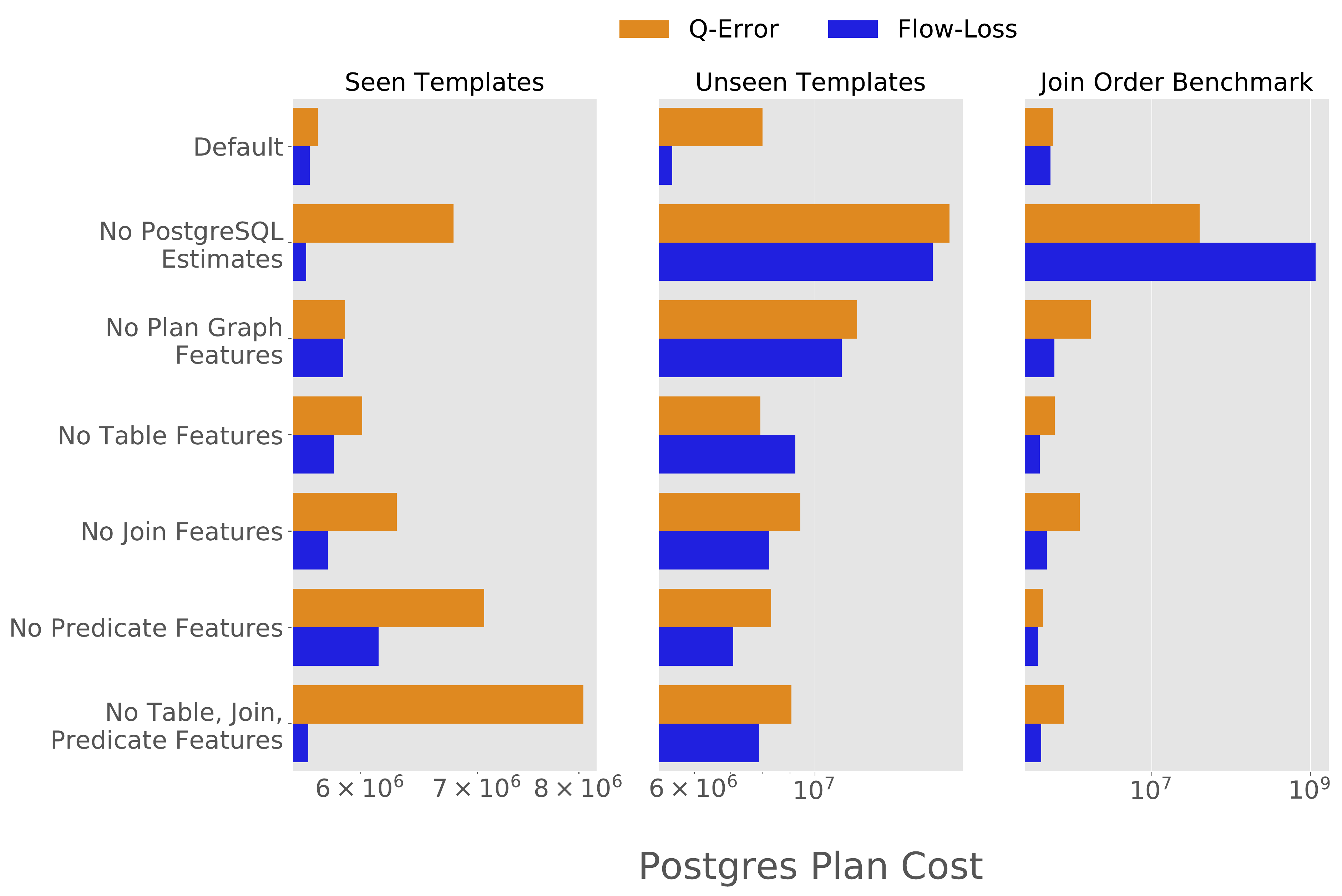}
\caption{{  Ablation study with the FCNN model for seen templates (left), unseen templates (middle), and JOB (right) showing \PPE when various components of the featurization scheme are removed. %
}}
\label{fig:ablation_study}
\end{figure}
\stitle{Ablation study.} Next, we seek to understand the impact of the various components of the featurization (cf. \S \ref{sec:featurization}) by an ablation study in which we remove key elements of the featurization, and evaluate the \PPE on the seen templates, unseen templates, and JOB.
We again focus on FCNN and omit MSCN, which follows similar trends.
Figure \ref{fig:ablation_study} summarizes the results. There are two main highlights. First, on the seen templates, \FlowLoss models can adapt to removing various featurization components, and do as well as with the default features, meanwhile, the Q-Error models suffer significantly with worse featurization. This shows that when constrained with fewer resources, the \FlowLoss model can better use its capacity to minimize \PPE. Second, PostgreSQL features are crucial for generalization. These include various cardinality and cost estimates (cf. \S \ref{sec:featurization}). Both the \FlowLoss and the Q-Error models get significantly worse on unseen templates without these features. This explains how the models do relatively well on different templates
--- these heuristic features have similar semantics across different kinds of queries. \PlanGraph features also seem to help more than others for generalization to unseen templates.

\stitle{Training time.} Compared to Q-Error, there is a $3-5\times$ overhead for training either architecture with \FlowLoss due to the additional calculations needed for \FlowLoss --- the bottleneck is computing $B(Y)^{-1}$ in Equation~\ref{eq:flows_linear_eq}.\footnote{
In the context of similar maximum flow algorithms on graphs, Christiano et al.~\cite{maxflow1} show ways to calculate $B(Y)^{-1}$ in polynomial time, utilizing the structure of the electric flows formulation.
We also expect it to be faster on GPUs with fast matrix inverse operations~\cite{matrixinverse}.
}
On the CPU, when using Q-Error, the FCNN architecture trains for $10$ epochs on the IMDb workload in under $1000$ seconds, and the MSCN model takes up to $2500$ seconds.

\stitle{Inference time.} As in~\cite{mscn,lightweightmodels}, the inference times for these neural networks is in the order of a few milliseconds (after featurization) and hence fast enough for query optimization.

\stitle{Model sizes.} %
The MSCN model is $2.4$MB, while the FCNN model is $4.7$MB. Sizes are the same for all loss functions.

\subsection{StackExchange workload}
\begin{figure}[t]
    \includegraphics[width=\columnwidth]{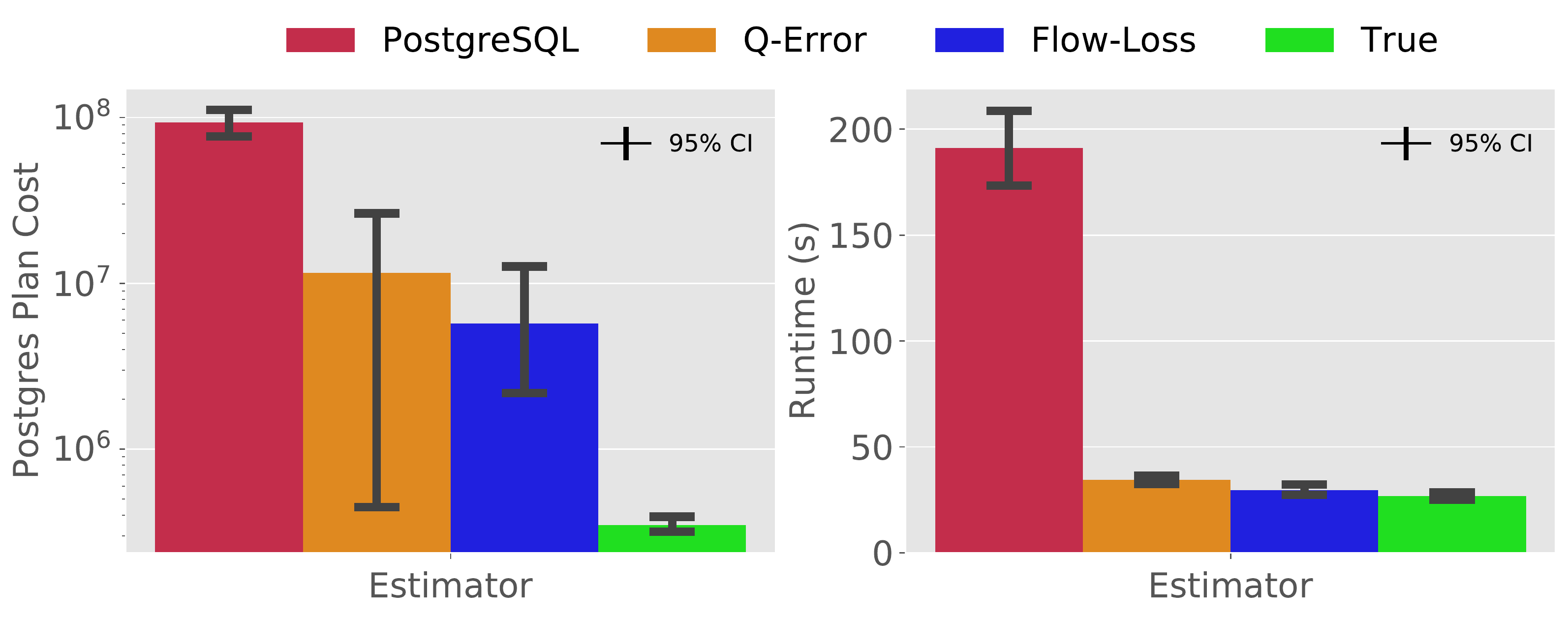}
    \caption[]%
    {{ Mean \PPE (five repeated runs) and runtimes (from one random run) for baselines and MSCN models evaluated on seen templates from the StackExchange workload.
    }}
    \label{fig:se_summary}
\end{figure}

Finally, we study the performance of the MSCN model on the StackExchange (SE) database using a workload that consists of fewer templates and queries than IMDb.

\stitle{Differences with IMDb results.} The SE workload has many timed out sub-plans which we replace with a large constant (cf. \S~\ref{sec:dataset}). A side effect is that it can artificially inflate the \PPE for plans that use such sub-plans because these large values would be used to compute the costs. Thus,  \PPE sees a lot more variance (cf. Figure~\ref{fig:se_summary}).

\stitle{Similarities to IMDb results.} 
On seen templates, both loss functions improve significantly over the PostgreSQL estimates, and perform similarly to each other, although the Q-Error models exhibit more variance on \PPE over five repeated experiments (cf. Figure \ref{fig:se_summary}).
As in the IMDb workload, sufficiently large improvements on \PPE translate to clear runtime improvements.
On unseen templates, the \FlowLoss models improve significantly over the Q-Error models (cf. Figure \ref{fig:meandiffs2}).
The magnitude of improvements are larger than on IMDb, with three partitions having improvements of over $20$ seconds on average. Partially, this is because the database size is also larger than IMDb --- thus better plans lead to more substantial improvements.

\section{Conclusions}
\label{sec:conclusion}
We showed that \PostgresError (\PPE) is a useful proxy to runtimes, and is an important alternative to Q-Error when evaluating a cardinality estimator. This lets us view cardinality estimation from a new lens --- and we developed \FlowLoss as a smooth, differentiable approximation to \PPE that can be used to train models via gradient descent based learning techniques. Using a new \OurDataset, we provide evidence that \FlowLoss can guide learned model's to better utilize their capacity to learn cardinalities that have the most impact on query performance. Even more importantly, it can help models avoid overfitting to cardinality estimates that are unlikely to improve query performance --- leading to more robust generalization when evaluated on queries from templates not seen in the training data, and helping models learn more robustly from training data generated using AQP techniques.
Generating ground truth cardinalities in order to train a model is expensive; moreover, updates to the data would quickly make such training data stale. 
Thus, avoiding the overhead of generating true labeled data can significantly improve adoption of learned cardinality estimation models in practice.

\clearpage

\bibliographystyle{abbrv}
\bibliography{andreas}

\clearpage
\appendix
\section{Metric based on Plan-Cost \label{app:plan_cost_metric}}

Definition \ref{def:plan_cost}, \S \ref{sec:definitions} gave the definition for P-Cost to evaluate the goodness of cardinality estimates. Here, we show that we can extend this to a pseudometric to compute the distance between any two cardinality vectors for a given query:

\begin{equation}
    \text{P-Error}(Y_1, Y_2) = |\text{P-Cost}(Y_1, \Y) -  \text{P-Cost}(Y_2, \Y)|
\end{equation}

Notice, we need to pre-compute the constant vector, $\Y$ for the given query. This is a pseudo-metric because it satisfies its three properties:

\begin{enumerate}
    \item $d(x,x) = 0$; Clearly, P-Error$(Y_1, Y_1) = 0$. Also, notice that there can be other points such that P-Error$(Y_1, Y_2) = 0$.
    \item $d(x,y) = d(y,x)$ (Symmetry); follows due to the absolute value sign in the definition of P-Error.
    \item $d(x,z) \leq d(x,y) + d(y,z)$ (Triangle Inequality); We will present the proof for this below.
\end{enumerate}

For notational convenience, we will use $P_e$ to refer to Plan-Error and $P_c$ to refer to Plan-Cost. $Y_t$ refers to $\Y$. Then the proof for the triangle inequality follows:

\begin{align}
\begin{split}
P_e(Y_1,Y_3) = |P_c(Y_1,Y_t) - P_c(Y_3, Y_t)| \\
 =  | P_c(Y_1, Y_t) - P_c(Y_2, Y_t)+ \\ P_c(Y_2, Y_t) - P_c(Y_3, Y_t) |  \\
\leq |  P_c(Y_1, Y_t) - P_c(Y_2, Y_t) | + \\ |  P_c(Y_2, Y_t) - P_c(Y_3,Y_t) | \\
= P_e(Y_1,Y2) + P_e(Y_2,Y_3)  \\
\end{split}
\end{align}

The first line is the definition of P-Error. In the second line, we are adding and subtracting same value $P_c(Y_2, Y_t)$. The third line follows from the definition of absolute value, which gives us exactly the statement of the triangle inequality that we were trying to prove.

\section{Flow-Loss Details \label{app:flow_loss}}

\subsection{Computing Flows \label{app:flow_loss_sol}}

\begin{figure}[ht]
\centering
\includegraphics[width=0.5\columnwidth]{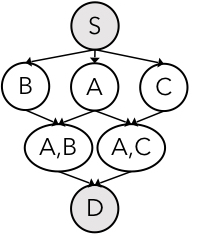}
\caption{{ Plan graph for query shown in Figure 1. }}
\label{fig:eg1_plan_graph}
\end{figure}
We will use the example of the query shown in Figure \ref{fig:eg1} in \S\ref{sec:intro} to show the explicit calculations used in the equations for computing the flows. First, we show the \plangraph for the query in Figure \ref{fig:eg1_plan_graph}. Our goal is to derive the solution for the optimization variable, $F$, in Equation \ref{eq:flows_optimization} in \S\ref{sec:fl}, which assigns a flow to every edge, $e$ in the \plangraph. Throughout this derivation, we will assume access to the cost function, $C$ (Definition \ref{def:cost_model}, \S \ref{sec:definitions}), which assigns $C(e,Y)$, a cost to any edge, $e$ in the \plangraph given a cardinality vector $Y$.
We rewrite the optimization program for $F$ below:

\begin{align}
\text{F-Opt(Y)} = \argmin_{F} \sum_{e \in E} C(Y)_e F_e^{2} \label{eq:flows_optimization-app} \\
\text{s.t} \sum_{e \in Out(S)} F_e = \sum_{e \in In(D)} F_e = 1 
\label{eq:opt_constraints1-app} \\
\sum_{e \in Out(V)} F_e = \sum_{e \in In(V)} F_e \label{eq:opt_constraints2-app}
\end{align}

For simplicity, we will use $F$ to refer to the vector solution of the optimization problem above.
We can express the set of constraints specified above as a system of linear equations. Recall, that in an electric circuit, every node has an associated voltage. Equation \ref{eq:flows_optimization-app} describes an electric circuit;
For our example (Figure \ref{fig:eg1_plan_graph}), we can write out the linear equations that need to be satisfied by the constraints in terms of the `voltages' of each node, and resistances (costs) for each edge, and using Ohm's law as:

\begin{align}
\begin{split}
    \frac{(v_S - v_A)}{C((S,A), Y)} + \frac{(v_S - v_B)}{C((S,B), Y)} + \frac{(v_S - v_C)}{C((S,C), Y)} &= 1 \\
    \frac{(v_B - v_{AB})}{C((B,AB), Y)} + \frac{(v_B - v_{S})}{C((S,B), Y)} &= 0 \\
    \frac{(v_C - v_{AC})}{C((C,AC), Y)} + \frac{(v_C - v_{S})}{C((S,C), Y)} &= 0 \\
    \frac{(v_A - v_{AB})}{C((A,AB), Y)} +  \frac{(v_A - v_{AC})}{C((A,AC), Y)} + \frac{(v_A - v_{S})}{C((S,A), Y)} &= 0 \\
    \frac{(v_{AB} - v_D)}{C((AB,D), Y)} +  \frac{(v_{AB} - v_A)}{C((A,AB), Y)} + 
    \frac{(v_{AB} - v_C)}{C((C,AB), Y)} &= 0 \\ 
    \frac{(v_{AC} - v_D)}{C((AC,D), Y)} +  \frac{(v_{AC} - v_A)}{C((A,AC), Y)} + \frac{(v_{AC} - v_C)}{C((C, AC), Y)} &= 0 \\ 
    \frac{(v_{D} - v_{AB})}{C((AB,D), Y)} +  \frac{(v_{AC} - v_A)}{C((A,AC), Y)} + \frac{(v_{AC} - v_C)}{C((C,AC), Y)} &= -1 \\ 
\end{split}
\label{eq:explicit_voltages}
\end{align}

where terms of the form $\frac{v_A}{C((S,A),Y)}$ use Ohm's law to compute the amount of maximum incoming current to node $A$ from $S$, and so on for the other terms.
Notice that the first and the last equations satisfy the constraint in Equation \ref{eq:opt_constraints1-app} --- $1$ unit of flow is outgoing from $S$ and $1$ unit of flow is incoming to $D$. All the intermediate equations represent the conservation constraint (Equation \ref{eq:opt_constraints2-app}) --- this will be obvious if you expand out each equations, and separate the terms being added (incoming current) versus the terms being subtracted (outgoing current) in each equation. 

Next, we will compactly represent the above linear constraints in terms of matrix operations. The system of linear equations above is over-determined, thus in practice, we remove the first equation and solve the remaining ones. For simplicity, we will express the matrix operations without removing any constraint.

Let $v \in R^{N}$ be the vector for the voltages $v_S ... v_D$ for all the nodes in the \plangraph. 
Similarly, let $i \in R^{N}$ be the vector of $1,0, ..., -1$ (the right hand side of Equation \ref{eq:explicit_voltages}). Let us define the edge-vertex incidence matrix, $X \in R^{n,m}$, with rows indexed by vertices and columns indexed by edges:

\begin{equation}
X_{n,e} = \begin{cases} 
      1 & \text{if  } e \in Out(n) \\
      -1 & \text{if  } e \in In(n) \\
      0 & \text{otherwise.}
   \end{cases}
   \label{eq:X}
\end{equation}

Next, for simplifying the notation, we will define $\frac{1}{C(e,Y)} = g_e$. And let, $G(Y) \in R^{m,m}$, be a diagonal matrix for the $m$ edges, with $G_{e,e} = g_e =  \frac{1}{C(e, Y)}$:

\begin{equation}
\begin{split}
g(Y) = 
    \begin{bmatrix} 
    \frac{1}{C(e_1,Y)} \\
    \vdots  \\
    \frac{1}{C(e_m,Y)}
    \end{bmatrix};
\hspace{2ex}
G(Y) = 
    \begin{bmatrix} 
    g(e_1)) & \dots & 0 \\
    \vdots & \ddots & \\
    0 & & g(e_m)
    \end{bmatrix}
\end{split}
\label{eq:G}
\end{equation}

Thus, $G$ is defined simply using the cost of each edge in the \plangraph w.r.t. a given cardinality vector $Y$. Note that in $g$ and $G$ we use the inverse of the costs, because in Equation \ref{eq:explicit_voltages}, the cost terms are in the denominator, thus this simplifies the formulas.

We define $B(Y) \in R^{N,N}$ as a linear function of the costs w.r.t. cardinality vector $Y$:
\begin{equation}
    B(Y) = X \cdot G(Y) \cdot X^{T}
    \label{eq:B}
\end{equation}

we can verify that each entry of $B$ is given by the following piece-wise linear function:
\allowbreak
\[B_{u,w} = \begin{cases} 
      \sum\limits_{e \in In(u) \cup Out(u)} \frac{1}{C(e,Y)} & \text{if  } u = w \\
      \allowbreak
      -\frac{1}{C(e,Y)} & \text{if  } e = (u,w) \text{ is an edge} \\
      
      0 & \text{otherwise.}
   \end{cases}
\]

Now, we are ready to compactly represent all the constraints from Equation \ref{eq:explicit_voltages}:
\begin{align}
\begin{split}
    B(Y) v = i \\
    \implies v = B(Y)^{-1}i
\end{split}
\end{align}

We do not know $v$, but $B(Y)$ is a deterministic function given a cost model, $C$ and cardinality vector $Y$, while $i$ is a fixed vector. Thus, using the constraints, we have found a way to compute the values for $V$. Note that $B(Y)$ does not have to be invertible, since we can use pseudo-inverses as well.

Recall, that the flows in Equation \ref{eq:flows_optimization-app} correspond to the current in the context of electrical circuits. Using Ohm's law, and the voltages calculated above, we can calculate the optimal flow (current) of an edge as:
\begin{equation}
 F_{(u,w)} = (v_u - v_w) \cdot g_{u,w}
\numberthis   
\end{equation}

where $g_{u,w} = \frac{1}{C((u,w), Y)}$ is the inverse of the cost of edge $(u,w)$, and $v_u, v_w$ are the voltages' associated with nodes $u$ and $w$. The linear equations for the flow, $F$, on each edge can be represented as a matrix multiplication:

\begin{align}
\begin{split}
F(Y) &= G(Y)  X  v \\
     &= G(Y)  X  B(Y)^{-1}i
\end{split}
\label{eq:flows_linear_solution}
\end{align}

where the second Equation uses Equation \ref{eq:flows_linear_solution}.
In \S\ref{sec:fl}, Equation \ref{eq:flows_linear_eq}, we gave the general form of the solution for $F(Y)$, which is satisfied by the precise definition in Equation \ref{eq:flows_linear_solution}.

\subsection{Flow-Loss}

Note that we found the flows, $F$, using the estimated cardinalities, $Y$, and the costs induced by them on each of the edges. To compute the true costs, which is used to calculate \FlowLoss, we will need to use the true cardinalities, $\Y$.  For convenience, we will define the following diagonal matrix with the true cost of each edge:
\begin{equation}
C^{true} = 
\begin{bmatrix} 
    C(e_1, \Y) & \dots & 0 \\
    \vdots & \ddots & \\
    0 & & C(e_m, \Y)
    \end{bmatrix}
\label{eq:Cstar}
\end{equation}

We will rewrite \FlowLoss, as also shows in Equation \ref{eq:flow_loss} in \S \ref{sec:fl}:
\begin{equation}
    \sum_e C_{e,e}^{true} F_e^2
    \label{eq:flow_loss-app}
\end{equation}

where we sum over all edges in the \plangraph. In terms of matrix operators, we can write it as: 
\begin{equation}
    F^T  C^{true} F
\end{equation}

\subsection{Flow-Loss gradient \label{app:flow_loss_grad}}

We present the dependency structure in the computation for $\FlowLoss$ below. Notice that the cardinalities, $Y$ only impact \FlowLoss through the costs, represented by $g$. 

$$
    Y \xrightarrow{C(\cdot)} \text{g} \xrightarrow{Opt(\cdot), C^{true}} \text{Flow-Loss}.
$$

For convenience, we use $g$ to represent costs --- recall, $g$ is just the inverse of the costs.
The definitions of $g$ and $C^*$ are given in Equations \ref{eq:G}, \ref{eq:Cstar}. %
For simplicity, we refer to $\hatY$ as simply $Y$.
We will use the vector notations $\vec{g}$ or $\vec{Y}$ to refer to vectors, and when gradients are taken w.r.t. vectors or scalar quantities. Recall, gradient of an $n$ dimensional vector w.r.t. an $m$ dimensional vector is the $nxm$ dimensional Jacobian matrix; while gradient of a scalar w.r.t. an $n$ dimensional vector is still an $n$ dimensional vector. For simplicity, we will also avoid explicitly writing out the dependencies as function arguments like we have done so far --- so instead of $G(Y)$ (Equation \ref{eq:G}), we will just write $G$.

\begin{align}
\begin{split}
    \nabla_{\vec{Y}} \text{\FlowLoss} = (\nabla_{\vec{Y}} \vec{g}) \; (\nabla_{\vec{g}} \text{\FlowLoss}) \\
            = (\nabla_{\vec{Y}} \vec{g}) \; (\nabla_{\vec{g}} \vec{F}) \;  (2\: C^{true} \vec{F})
\end{split}
\label{eq:gradient_basic_formula}
\end{align}

where $\vec{F}$ are the flows for each edge, which we can compute using Equation \ref{eq:flows_linear_solution}. The first line follows because given $g$ (inverse estimated costs of each edge), computing the \FlowLoss does not depend on $Y$. Therefore, we can use the chain rule to separate it out into two independent gradients. The second line follows because consider the partial derivative of \FlowLoss (as in Equation \ref{eq:flow_loss-app}) w.r.t. a single element of $g$:
\begin{align}
    \begin{split}
        \frac{\partial \text{\FlowLoss}}{\partial g_j} = 2 \sum_{e} C^{true}_{e,e} F_e \frac{\partial F_e}{\partial g_j}
    \end{split}
\end{align}

Here, $2\: \sum_{e} C^{true}_{e,e} F_e$ corresponds to the term $(2C^{true} F)$ in Equation \ref{eq:gradient_basic_formula}.

Next, we will write out the explicit formulas for each of the unknown terms in the above equation. First, $\nabla_{\vec{Y}}\vec{g}$ is the Jacobian matrix of the cost function, with input estimated cardinalities, and outputs costs.

\begin{equation}
\nabla_{\vec{Y}} \vec{g} = 
\begin{bmatrix} 
    \frac{dg_1}{dY_1} & \frac{dg_2}{dY_1} & \dots \frac{dg_m}{dY_1} \\
    \vdots & \ddots & \\
    \frac{dg_1}{dY_n} & & \frac{dg_m}{dY_n}
    \end{bmatrix}
\end{equation}

Notice that this is the only place where we need to take the gradient of the cost function, $C$ (Definition \ref{def:cost_model}, \S \ref{sec:definitions}). Thus, we just need to know how to take the gradient of the cost of a single edge w.r.t. the two cardinality estimates that are used to compute that cost, as in Definition \ref{def:cost_model}, \S \ref{sec:definitions}. Thus, we could potentially be using significantly more complex cost models as long as this value can be approximated. 
Also, most terms in this Jacobian matrix end of trivially being zeros since the cost of a particular edge will only depend on two elements of $Y$. This is one of the ways we can significantly speed up the gradient computations by explicitly coding up the formulas.

Calculating $\nabla_{\vec{g}} \vec{F}$ is more involved. The solution comes out to be:
\begin{equation}
\nabla_{\vec{g}} \vec{F} \in R^{M,N} = 
    \begin{bmatrix} 
    i^{T} B^{-T} ( \frac{\partial{GX}}{\partial g_1} - GXB^{-1} \frac{\partial{ G}}{\partial g_1} )  \\
    \vdots \\
    i^{T} B^{-T} ( \frac{\partial{GX}}{\partial g_1} - GXB^{-1} \frac{\partial{ G}}{\partial g_m} )  \\
    \end{bmatrix}
\end{equation}

Note, each row in the above matrix is a vector in $R^n$. $X$ was defined in Equation \ref{eq:X}, $B$ was defined in Equation \ref{eq:B}, and $G$ was defined in Equation \ref{eq:G}. As we can see there are many results being re-used from the computations of $F$; in terms of implementation, this means that the forward and backward passes of a neural network can reuse intermediate results, which results in significantly more efficient code as well. 
Also, once again we see that many of the partial derivatives would be $0$, which don't need to be computed when implementing these gradients.

\section{Dataset \label{app:dataset}}

\subsection{StackExchange timeouts \label{app:se_timeouts}}

\begin{figure}[ht]
\centering
\includegraphics[width=0.75\columnwidth]{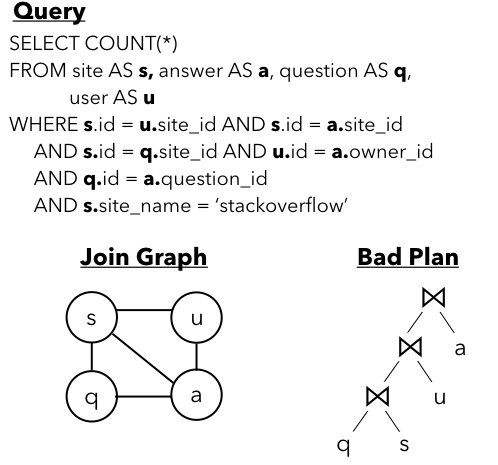}
\caption{An example of a timed out subquery on the StackExchange database.}
\label{fig:stackexchange_timeout}
\end{figure}

On the stackexchange datasets, three of the six templates have a large proportion of timeouts when generating the ground truth data for the sub-plans. This is due to the unusual join graph, presented with a simplified query in Figure \ref{fig:stackexchange_timeout}. 
As the accompanying query makes clear, there is a relationship between the joins on user, $u$ and answer $a$. But there is no relationship between question $q$ and user $u$ without $a$; still, the sub-plan $q \Join s \Join u$ is not a cross-join since all tables have a relationship with site $s$. Essentially, $q \Join s \Join u$ will give us the cross-join of all questions (potentially, satisfying some filters) along with all users (satisfying some filters) --- which will blow up and lead to timeouts. Such join graphs can occur in natural queries, for instance, to process information about users who answer certain types of questions (e.g., what is the location of users who answer stackoverflow questions with Javascript tags).

\end{document}